\documentclass[pdflatex,sn-mathphys]{sn-jnl}
\catcode`\|=12\relax 

\usepackage{graphicx}
\usepackage{hyperref}
\usepackage{subcaption}
\usepackage{float}
\usepackage[capitalize]{cleveref}
\usepackage{adjustbox}
\usepackage{xcolor}
\usepackage{bm}
\usepackage[normalem]{ulem}

\graphicspath{{figs/}}

\begin{document}

\title[Rapid training of quantum recurrent neural networks]{Rapid
training of quantum recurrent \\neural networks}
\date{\today}

\author*[1]{\fnm{Micha\l} \sur{Siemaszko}}
\email{mm.siemaszko2@uw.edu.pl}

\author[1]{\fnm{Adam} \sur{Buraczewski}}

\author[2]{\fnm{Bertrand} \sur{Le~Saux}}

\author[1]{\fnm{Magdalena} \sur{Stobi\'nska}}

\affil*[1]{\orgdiv{Faculty of Mathematics, Informatics and Mechanics},
\orgname{University of Warsaw}, \country{Poland}}

\affil[2]{\orgdiv{$\Phi$-lab}, \orgname{European Space Agency},
\city{Frascati}, \country{Italy}}

\abstract{Time series prediction is essential for human activities in
  diverse areas. A common approach to this task is to harness Recurrent
  Neural Networks (RNNs). However, while their predictions are quite
  accurate, their learning process is complex and, thus, time and
  energy consuming. Here, we propose to extend the concept of RRNs by
  including continuous-variable quantum resources in it, and to use a
  quantum-enhanced RNN to overcome these obstacles. The design of the
  Continuous-Variable Quantum RNN (CV-QRNN) is rooted in the
  continuous-variable quantum computing paradigm. By performing extensive
  numerical simulations, we demonstrate that the quantum network is
  capable of learning-time dependence of several types of temporal data,
  and that it converges to the optimal weights in fewer epochs than a
  classical network. Furthermore, for a small number of trainable
  parameters, it can achieve lower losses than its classical counterpart.
CV-QRNN can be implemented using commercially available quantum-photonic
hardware.}


\maketitle

\clearpage


\section{Introduction}

Fast and accurate time series analysis and prediction lie at the heart of
digital signal processing, and machine learning algorithms help implement
them~\cite{gamboaDeepLearningTimeSeries2017,limTimeSeriesForecasting2021}.
They feature a wide palette of use cases ranging from audio and video
signal processing and compression~\cite{maImageVideoCompression2020},
temporal signal classification~\cite{huskenRecurrentNeuralNetworks2003},
speech processing and recognition~\cite{amodeiDeepSpeechEndtoEnd2016,
dahlContextDependentPreTrainedDeep2012, sakLongShortTermMemory2014},
economic~\cite{saadComparativeStudyStock1998} and earth system
observation \cite{bonavitaMachineLearningEarth2021,
holmstromMachineLearningApplied2016}, to applications in
seismology~\cite{kongMachineLearningSeismology2018} and
biomedicine~\cite{goecksHowMachineLearning2020}.  A method that is
particularly well suited for the analysis of temporal correlations in
data sequences is Recurrent Neural Networks
(RNNs)~\cite{sherstinskyFundamentalsRecurrentNeural2020}. This is because
they accumulate information about subsequent input data, which amounts to
a cumulative memory effect seen in their computations.  However, RNNs
often suffer from a vanishing or exploding gradient, which hampers the
training process and makes it energy and time
consuming~\cite{bengioLearningLongtermDependencies1994,
pascanuDifficultyTrainingRecurrent2013}.  Although efforts have been made
to overcome this issue by harnessing, e.g.\ the Long Short-Term Memory
(LSTM)~\cite{hochreiterLongShorttermMemory1997}, the Gated Recurrent
Units (GRU)~\cite{choPropertiesNeuralMachine2014} or Unitary Recurrent
Neural Networks (uRNN) \cite{arjovskyUnitaryEvolutionRecurrent2016,
wisdomFullCapacityUnitaryRecurrent2016}, the problem remains open. This
motivates us to continue developing these platforms by including quantum
resources.

Quantum Machine Learning (QML)
\cite{schuldIntroductionQuantumMachine2014,
biamonteQuantumMachineLearning2017} holds promise of augmenting the
machine learning process by employing quantum resources and speeding up
computations. To this end, both qubit (discrete-variable, DV) and
continuous-variable (CV) data encodings are extensively studied~\cite{
gargAdvancesQuantumDeep2020, garciaSystematicLiteratureReview2022}.  CV
systems can be implemented with quantum-photonic platforms and trapped
ions.  Acceleration of computations can be achieved either by lowering
the computational complexity of an algorithm by its quantum
implementation~\cite{rebentrostQuantumSupportVector2014,
schuldPredictionLinearRegression2016, liuRigorousRobustQuantum2021} or by
reducing the time of learning process, i.e.\ the number of epochs
required for a NN to complete it. The latter approach is at the focus of
this work. It is usually pursued by means of parameterized quantum
circuits~\cite{schuldEffectDataEncoding2021,
  farhiClassificationQuantumNeural2018,
schuldCircuitcentricQuantumClassifiers2020,
benedettiParameterizedQuantumCircuits2019} where the values of quantum
gates' parameters come as a result of circuit training. This recursive
process is similar in spirit to the Feed Forward Neural Network
algorithm~\cite{svozilIntroductionMultilayerFeedforward1997}. This method
has recently been proven to be useful for satellite image
classification~\cite{sebastianelliCircuitBasedHybridQuantum2022}, joint
probability distribution
modelling~\cite{zhuGenerativeQuantumLearning2022}, and time series
analysis~\cite{bauschRecurrentQuantumNeural2020,
  takakiLearningTemporalData2021, chenQuantumLongShortTerm2022,
emmanoulopoulosQuantumMachineLearning2022}.

Until now, quantum-enhanced implementations of RNNs that have been used
for time series analysis, were designed for multiple-qubit data input.
One such quantum modification of RNNs is the Recurrent Quantum Neural
Network (RQNN)~\cite{bauschRecurrentQuantumNeural2020}. In this network
each cell is built from a parametrized neuron and amplitude amplification
serves as a nonlinear function applied after each cell call. On the
contrary, the Quantum Recurrent Neural Network
(QRNN)~\cite{takakiLearningTemporalData2021} consists of cells made of
parametrized quantum circuits, which are capable of performing unitary
transformations on all input qubits. An alternative approach to temporal
data prediction is based on Quantum Long Short-Term Memory
(QLSTM)~\cite{chenQuantumLongShortTerm2022}. It employs a classical
architecture, in which LSTM cells are replaced with parametrized quantum
circuits optimized during the training process. The idea of constructing
the Quantum Gated Recurrent Unit (QGRU) was proposed and analyzed in
\cite{chenQuantumRecurrentEncoder2020}. In the last two cases, the
implementation was based on internal measurements of the quantum state to
realize necessary additional operations and rule sets, which rendered
these approaches semi-classical.  A different variant of the quantum
recurrent neural network was proposed in
\cite{hibat-allahRecurrentNeuralNetwork2020}, where a variational
wave-functions were used to learn the approximate ground state of a
quantum Hamiltonian. Finally, the Hopfield Network, which is a form of an
RNN, has awaited several implementations on a quantum computer
\cite{rebentrostQuantumHopfieldNeural2018,
rotondoOpenQuantumGeneralisation2018,
tangExperimentalQuantumStochastic2019}
 
Here, we propose a RNN-based quantum algorithm for rapid and rigorous
analysis and prediction of temporal data in the CV regime (CV-QRNN).
CV-QRNN capitalizes on the parameterized quantum circuit proposed
in~\cite{killoranContinuousvariableQuantumNeural2019}. Its operation
cycle consists of three phases: entering data, processing them, and
performing a measurement. The measurement result, together with the next
data point, constitutes the input for the next cycle. To the best of our
knowledge, we are the first to construct and study a QRNN in the CV
regime for time series processing. We train CV-QRNN for sequence data
prediction, forecasting, and image classification and compare the results
with the state-of-the-art LSTM implementation. By means of extensive
numerical simulations, we demonstrate significant reduction of the number
of epochs required for CV-QRNN training to achieve similar results
compared to a fully classical implementation with a comparable number of
tunable parameters.

This paper is organized as follows. \Cref{sec:background} describes
CV-QRNN's theoretical model and its architecture. In~\Cref{sec:numerical}
we demonstrate results of our numerical simulations, with the methods
described in \Cref{sec:numerical_methods}. The conclusions and discussion
are provided in~\Cref{sec:results}.


\section{Theoretical model}
\label{sec:background}


\subsection{Continuous-variable quantum information processing}
\label{sec:cv}

There are two main quantum information frameworks explored. In one of
them, information is encoded in discrete variables that are represented
by qubits, and in the other one in continuous variables, embodied by
qumodes. Both schemes facilitate universal quantum computation, i.e.\
they can implement an arbitrary unitary evolution with arbitrarily small
error~\cite{weedbrookGaussianQuantumInformation2012,
lloydQuantumComputationContinuous1999}.  While qubits are a counterpart
of classical digital computation with bits, CVs resemble analog
computing. Here we focus on the CV quantum framework.

Quantum CV systems hold promise of performing computations more
effectively than their DV
counterparts~\cite{lloydQuantumComputationContinuous1999}. In particular,
thanks to the ability of CV systems to deterministically prepare large
resource states and to measure results with high efficiency using
homodyne detection, they scale up
easily~\cite{guQuantumComputingContinuousvariable2009}, leading e.g.\ to
instantaneous quantum computing
(IQP)~\cite{douceContinuousVariableInstantaneousQuantum2017}. These
hypothesis is also reinforced by the fact that classical analog
computation has been shown to be effective in solving differential
equations, some optimization problems, and simulations of nonlinear
physical systems~\cite{vergisComplexityAnalogComputation1986}, where it
is able to achieve accurate results in a very short
time~\cite{chuaNonlinearProgrammingComputation1984}. Analog accelerators
have been proposed as an efficient implementation of deep neural
networks~\cite{xiaoAnalogArchitecturesNeural2020}. 

Universal CV quantum computation requires a set of single-qumode gates
and one controlled two-qumode gate that will generate all possible
Gaussian operations, as well as one single-qumode nonlinear
transformation of polynomial degree 3 or
higher~\cite{lloydQuantumComputationContinuous1999,
weedbrookGaussianQuantumInformation2012}. In the case of quantum photonic
circuits, qumodes are realized by photonic modes that carry information
encoded in the quadratures of the electromagnetic field. These
quadratures possess a continuous spectrum and constitute the CVs with
which we compute. All Gaussian gates can be built from simple linear
devices such as beam splitters, phase shifters, and
squeezers~\cite{knillSchemeEfficientQuantum2001}.  Nonlinearity is
usually achieved by cross-Kerr
interaction~\cite{stobinskaWignerFunctionEvolution2008},
but it can also be induced by the measurement process, either
photon-number-resolving~\cite{scheelMeasurementinducedNonlinearityLinear2003}
or homodyne~\cite{filipMeasurementinducedContinuousvariableQuantum2005}.

The implementation of CV-QRNN will involve the displacement gate 
\begin{equation}
    D(\alpha) := 
    \exp\bigl\{ \alpha\hat{a}^\dagger - \alpha^{*}\hat{a} \bigr\},
\end{equation}
where $\alpha$ is a complex displacement parameter, $\hat{a}$
($\hat{a}^\dagger$) is a qumode annihilation (creation) operator,
respectively. We will also use the squeezing gate
\begin{equation}
    S(r) := 
    \exp\bigl\{ \tfrac{r}{2}(\hat{a}^2 + \hat{a}^{\dagger2})\bigr\},
\end{equation}
where $r$ is a complex squeezing parameter, as well as the phase gate
\begin{equation}
    R(\varphi) := \exp\bigl\{ -i\varphi\hat{a}^\dagger \hat{a} \bigr\},
\end{equation}
with phase $\varphi \in (0,2\pi)$. We will also harness the beam splitter
gate, which is the simplest two-input and two-output interferometer,
\begin{equation}
    B(\theta) := \exp\bigl\{ 
    \theta (\hat{a}^\dagger\hat{b} - \hat{a}\hat{b}^\dagger) \bigr\},
\end{equation}
where $\theta\in (0,\tfrac{\pi}{2})$, $\hat{a}$ and $\hat{b}$
($\hat{a}^\dagger$ and $\hat{b}^\dagger$) are annihilation (creation)
operators of two interfering qumodes, respectively. Any arbitrary
multiport interferometer, denoted here by $I(\bm{\theta}, \bm{\varphi})$,
can be implemented with a network of phase and beam splitter
gates~\cite{reckExperimentalRealizationAny1994}. In our work, we will use
the Clemets decomposition \cite{clementsOptimalDesignUniversal2016} to
achieve this goal. All described gates are implementable with the
commercially available quantum-photonic hardware. To realize nonlinear
operations, CV-QRNN will harness the tensor product structure of a
quantum system~\cite{zanardiQuantumTensorProduct2004}, which is capable
of providing nonlinearity by means of measurement, in the spirit of
Refs.~\cite{killoranContinuousvariableQuantumNeural2019,
takakiLearningTemporalData2021}.  This will free us from the necessity of
utilizing strong Kerr-type interactions that are difficult to implement.


\subsection{Recurrent Neural Networks}
\label{sec:rnn_architecture}

\begin{figure}
    \centering
    \includegraphics[width=0.9\textwidth]{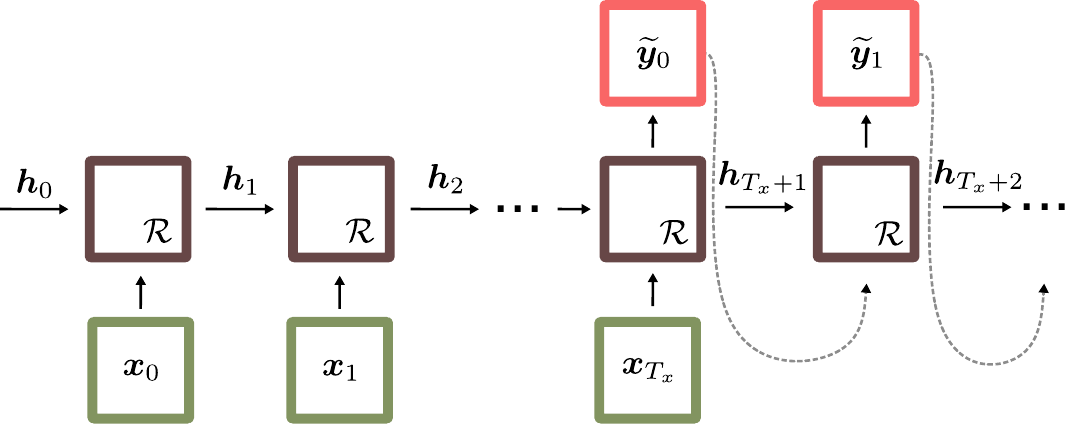}

    \vspace{0.5cm}

    \caption{Schema of a Recurrent Neural Network. At every time step
        $t$, an input vector $\bm{x}_t$ is injected to the network cell
        (brown square) that is parametrized by a hidden state $\bm{h}_t$.
        After all the input data have been processed, output sequences
        $\widetilde{\bm{y}}_\tau$ are produced and they serve as the next
        input to the RNN (dashed arrows). Parameters of the network (not
        shown on the figure) are described in the text. Additional sets
        of rules $\mathcal{R}$ included in the network cells upgrade RNN
    to LSTM or GRU architectures.}
    \label{fig:rnn}
\end{figure}

Our quantum-enhanced RNN architecture (CV-QRNN) is inspired by the
vanilla RNN depicted in
\cref{fig:rnn}~\cite{sherstinskyFundamentalsRecurrentNeural2020}. This is
a standard network layout which is trained by iterating over the elements
of an input data sequence. Then, during the prediction phase, the output
values are looped back to the input to obtain subsequent results.

In the RNN, $T_x$ $n$-bit input sequences $\{\bm{x}_i\}_{i=0}^{T_x}$
($\bm{x}_i \in \mathbb{R}^n$, indicated as green squares in
\cref{fig:rnn}) are sequentially processed by a cell (brown square) to
produce $T_y$ $m$-bit output sequences
$\{\widetilde{\bm{y}}_i\}_{i=0}^{T_y}$ ($\widetilde{\bm{y}}_i \in
\mathbb{R}^m$, pink squares).  At each time step $t$, the RNN cell is
characterized by a \textit{hidden state} vector $\bm{h}_t \in
\mathbb{R}^d$, which serves as a memory that keeps the internal state of
the network. It is updated as soon as a new data point is injected into
the network in step $t+1$
\begin{equation} 
    \bm{h}_{t+1} = 
    \begin{cases}
        g_h(W_x \bm{x}_{t} + W_h \bm{h}_{t} + \bm{b}_h), &  
        0 \le t \le T_x\\
        g_h(W_x \widetilde{\bm{y}}_{t-T_x-1} + 
        W_h \bm{h}_{t} + \bm{b}_h), &
        t > T_x
    \end{cases}
\end{equation}
where $W_x, W_h$ are weight matrices of dimensions $d\times n$ and $d
\times d$, respectively, $\bm{b}_h \in \mathbb{R}^d$ is a bias vector,
$g_h$ is an element-wise nonlinear activation function.  $\bm{h}_0$ is an
initial hidden state which is a parameter of the network.

The output sequences are computed only after all input data points were
processed by the RNN
\begin{equation}
    \widetilde{\bm{y}}_\tau = 
    g_o\left( W_y \bm{h}_{T_x+\tau} + \bm{b}_y \right),
\end{equation}
where $W_y$ is a weight matrix of dimension $m \times d$, $\bm{b}_y \in
\mathbb{R}^m$ is a bias vector and $g_o$ is an element-wise nonlinear
activation function, which can be different from $g_h$. 

Next, we validate the accuracy of the results produced by the network. To
this end, we compute a cost function $C$ that allows us to compare
$\{\widetilde{\bm{y}}_t\}_{t=0}^{T_y}$ with the desired result
$\{\bm{y}_t\}_{t=0}^{T_y}$. In the case of the sequence prediction and
forecasting task, the mean square error was adopted
\begin{equation}
    C_{MSE}\left(\{\widetilde{\bm{y}}_t\}_{t=0}^{T_y},
    \{\bm{y}_t\}_{t=0}^{T_y}\right) = \frac{1}{m} \sum_{t=0}^{T_y} \|
    \widetilde{\bm{y}}_t - \bm{y}_t\|^2,
    \label{eq:cost}
\end{equation}
while for the classification task -- the binary cross entropy, in which
only a single output $\widetilde{\bm{y}}_0 \equiv \widetilde{\bm{y}}$ is
compared to the expected label $\bm{y}_0 \equiv \bm{y}$
\begin{equation}
    C_{BCE}\left( \widetilde{\bm{y}}, \bm{y}\right) = 
    \frac{1}{m} \sum_{i=1}^m - \left( y_i \log(\widetilde{y}_i) + 
    (1-y_i) \log(1 - \widetilde{y}_i) \right).
    \label{eq:cost2}
\end{equation}
Minimization of the cost function by means of backpropagation helps us to
optimize parameters of the network.  The state-of-the-art LSTM and GRU
architectures introduce a modification to RNNs by complementing the
hidden layer with additional sets of rules $\mathcal{R}$ that determine
how long the information about previous data points should be
kept~\cite{hochreiterLongShorttermMemory1997}. It is implemented by
functions acting on copies of input and hidden layer data, which amplify
or vanish selected values from previous iterations. We use LSTM as a
classical reference system to which we compare the performance of
CV-QRNN. We find this comparison fair because LSTM is one of the most
widely used schemes in industrial
applications~\cite{vanhoudtReviewLongShortterm2020} that is similar in
its architecture and mode of operation to CV-QRNN. In this paper, we use
its implementation, which follows the original proposal found in
Ref.~\cite{hochreiterLongShorttermMemory1997}.


\subsection{CV-QRNN architecture}
\label{sec:architecture}

\begin{figure}
    \centering
    \begin{subfigure}{\textwidth}
    \begin{center}
        \includegraphics{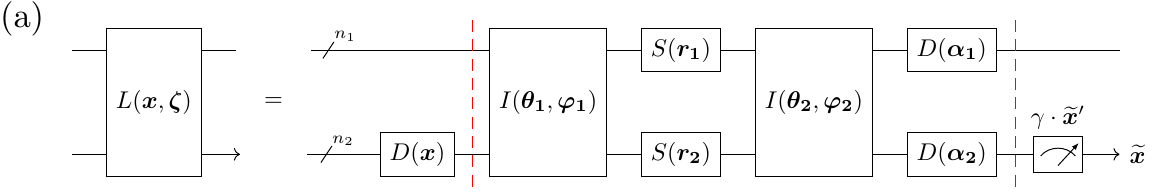} 
    \end{center}
    \phantomsubcaption\label{fig:circuit_a}
    \end{subfigure}

    \vspace{0.2cm}

    \begin{subfigure}{\textwidth}
    \begin{center}
        \includegraphics{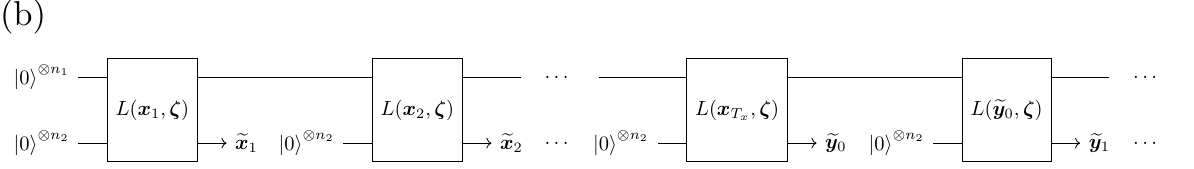} 
    \end{center}
    \phantomsubcaption\label{fig:circuit_b}
    \end{subfigure}

    \vspace{0.7cm}

    \caption{CV-QRNN architecture. (a) Single layer $L$ acts on $n = n_1
        + n_2$ qumodes (horizontal lines), and consists of displacement
        gates $D$, squeezing gates $S$, and multiport interferometers
        $I$. A vector $\bm{x} \in \mathbb{R}^{n_2}$ encodes the input
        data, while $\bm{\zeta} = \{\bm{\theta_1}, \bm{\varphi_1},
            \bm{r_1}, \bm{r_2},\bm{\theta_2}, \bm{\varphi_2},
        \bm{\alpha_1},\bm{\alpha_2}, \gamma\}$ denotes all trainable
        parameters of the network. Red dashed lines split the layer into
        three parts, responsible for (from left to right): encoding,
        interaction, and measurement.  (b) Data sequence is processed
        recurrently by iterating layer $L$ over all inputs
        $\bm{x_1},\ldots,\bm{x}_{T_x}$. All the qumodes are initialized
        with the vacuum state $\vert 0 \rangle^{\otimes n_{1,2}}$.  After
        each iteration, the output $\widetilde{\bm{x}}'_t$ is measured,
        mulitplied by parameter $\gamma$, and all bottom wires are reset
        to the vacuum state. The first prediction of the network
        $\widetilde{\bm{y}}_0$ is taken only after all data points have
        been processed. The subsequent prediction
        $\widetilde{\bm{y}}_\tau$ is the output of the layer  $L\left(
    \widetilde{\bm{y}}_{\tau-1}, \bm{\zeta} \right) $}
    \label{fig:circuit} 
\end{figure}

The detailed CV-QRNN layout, shown in \cref{fig:circuit}, is based on a
vanilla RNN. This is because GRU and LSTM architectures cannot be
directly implemented on a quantum computer as a result of the no-cloning
theorem (the no-cloning theorem forbids to copy quantum information). In
addition, quantum memories, which are required to implement internal
rules in the latter networks, are unfeasible.

The wires represent the $n$-dimensional tensor product of the qumodes,
and the rectangles represent the quantum gates. Each qumode is initially
prepared in the vacuum state $\lvert0\rangle$, which is collectively
denoted as $\lvert 0\rangle^{\otimes n}$. To highlight the fact that
every gate acts on $n$ qumodes simultaneously, but each qumode sees
different gate parameters, we use the following notation: $D(\bm{v})
\equiv \bigotimes_i D(v_i)$ and $S(\bm{v}) \equiv \bigotimes_i S(v_i)$,
where $\bm{v} = \left( v_1,\ldots,v_n \right)^{\text{T}}$, $\bigotimes$
is the tensor product, $D$ and $S$ are a single-qumode displacement and
squeezing gates, respectively.

A single quantum layer $L$, shown in \cref{fig:circuit_a}, acts in the
following way: first, it encodes classical data $\bm{x}$ into the quantum
network by means of a displacement gate $D(\bm{x})$ that acts on $n_2$
qumodes prepared in the vacuum state $\vert 0 \rangle^{\otimes n_{2}}$
(bottom wire).  Next, all $n=n_1+n_2$ qumodes (top and bottom wires) are
processed in a multiport interferometer $I(\bm{\theta_1},
\bm{\varphi_1})$ followed by squeezing gates $S(\bm{r_{1,2}})$, another
interferometer $I(\bm{\theta_2}, \bm{\varphi_2})$, and displacement gates
$D(\bm{\alpha_{1,2}})$. As a result of this, the layer $L$ outputs a
highly entangled state that involves all $n$ qumodes. Eventually, $n_2$
qumodes are subjected to a homodyne measurement and reset to the vacuum
state, while $n_1$ qumodes are passed to the next iteration. 

The qumodes that are measured are dubbed the \textit{input modes}, while
these left untouched -- the \textit{register modes}. The output of the
former, $\widetilde{\bm{x}}$, equals to the mean value of the measurement
results $\widetilde{\bm{x}}'$ multiplied by the trainable parameter
$\gamma$. For convenience of notation, we denote all the gates'
parameters in the network as $\bm{\zeta} = \{ \bm{\theta_1},
    \bm{\varphi_1}, \bm{r_1}, \bm{r_2},\bm{\theta_2}, \bm{\varphi_2},
\bm{\alpha_1},\bm{\alpha_2}, \gamma\}$. Thus, the layer $L$ is
characterized by $2\left( n^2 + \max(1,n-1) \right) + n + 1$ parameters
in total, which are randomly initialized before the first run.

Sequential processing of data points $\{\bm{x}_i\}_{i=0}^{T_x}$ is shown
in \cref{fig:circuit_b}.  As soon as the quantum layer $L(\bm{x}_t,
\bm{\zeta})$  is executed in the time step $t$, the bottom $n_2$ qumodes
are reset to the vacuum state $\vert 0 \rangle^{\otimes n_2}$ and fed to
the next layer $L(\bm{x}_{t+1},\bm{\zeta})$ along with $n_1$ qumodes that
were never measured. This process is iterated $T$ times. The data point
that follows $\bm{x}_{T_x}$ is $\widetilde{\bm{x}}_{T_x} \equiv
\widetilde{\bm{y}}_0$ and the process continues, i.e.\ the layer
$L(\widetilde{\bm{y}}_{\tau}, \bm{\zeta})$ outputs
$\widetilde{\bm{y}}_{\tau+1}$, for the next $T_y$ steps.  Only the output
$\bm{y}_0,\ldots,\bm{y}_{T_y}$ is then analyzed.


\section{Numerical simulations}
\label{sec:numerical}

To assess the quantum-enhanced performance of the CV-QRNN architecture
depicted in \cref{fig:circuit}, we compared its performance with a
classical LSTM (\cref{fig:rnn}). Our figure of merit was the reduction in
the number of epochs required to obtain a clear plateau in subsequent
values of the cost function $C$, which achieve the same order of
magnitude as for the reference classical network. The comparison involved
running the quantum algorithm under a software simulator of a CV quantum
computer, which was used to calculate the measurement outputs of the
layer $L$ and optimize the trainable parameters $\bm{\zeta}$. Reference
data were obtained by processing the same input with a state-of-the-art
LSTM implementation. For our experiments, we chose two tasks to be
realized by both networks: time series prediction and forecasting, as
well as data classification.  The former demonstrated the ability of
CV-QRNN architecture to compute subsequent data values from initial
samples of periodic or quasi-periodic functions. The latter was a
textbook classification problem of recognizing MNIST handwritten digits
based on the initial learning of the network. It allowed us to show that
even a small number of parameters was suitable for correct discrimination
between data sets.

\textbf{Task 1 -- sequence prediction and forecasting.}  We define
prediction as computing only a single value of the function $f(x)$ based
on the previous $T$ data points in a sequence; and forecasting as
computing several consecutive values to achieve a longer output. For this
task, we chose quasi-periodic Bessel function of degree 0, $f(x)\equiv
J_0(x)$. It has wide applications in physics and engineering, as it
describes various natural
processes~\cite{korenevBesselFunctionsTheir2002}. Since the oscillation
amplitude vanishes for large $x$, forecasting of this function is
non-trivial. The Bessel function was used to generate 200 equidistant
points $\bigl(x_i,J_0(x_i)\bigr)$, where $x_0=0$ and $x_{200} \approx
4\Omega$ and $\Omega$ designates the function period. Next, taking
$\overline{x}_i = J_0(x_i)$, we computed the sequence
$\{\overline{x}_i\}_{i=0}^{200}$. It was split equally between the
training and test data sets, so that each set contained 2 periods of the
function. The network was trained to predict $\overline{x}_{i+T}$ based
on the input that consisted of $T-1$ previous data points. For each input
$\{\overline{x}_i, \ldots, \overline{x}_{i+T-1}\} $, where
$i=0,\ldots,200-T$, the network returned the output $y$, which was
trained to be as close to $x_{i+T}$ as possible. We used $T=4$ (the
rationale for this choice is presented below).  The standard baseline
model for this task was to repeat the last input point as the output
value, $\overline{x}_{i+T} = \overline{x}_{i+T-1}$. The results achieved
for other functions, such as sine, triangle wave, and damped cosine, are
shown in the Appendix~\ref{sec:additional_datasets}.

\begin{figure}[ht]
    \centering
    \includegraphics[width=0.8\textwidth]{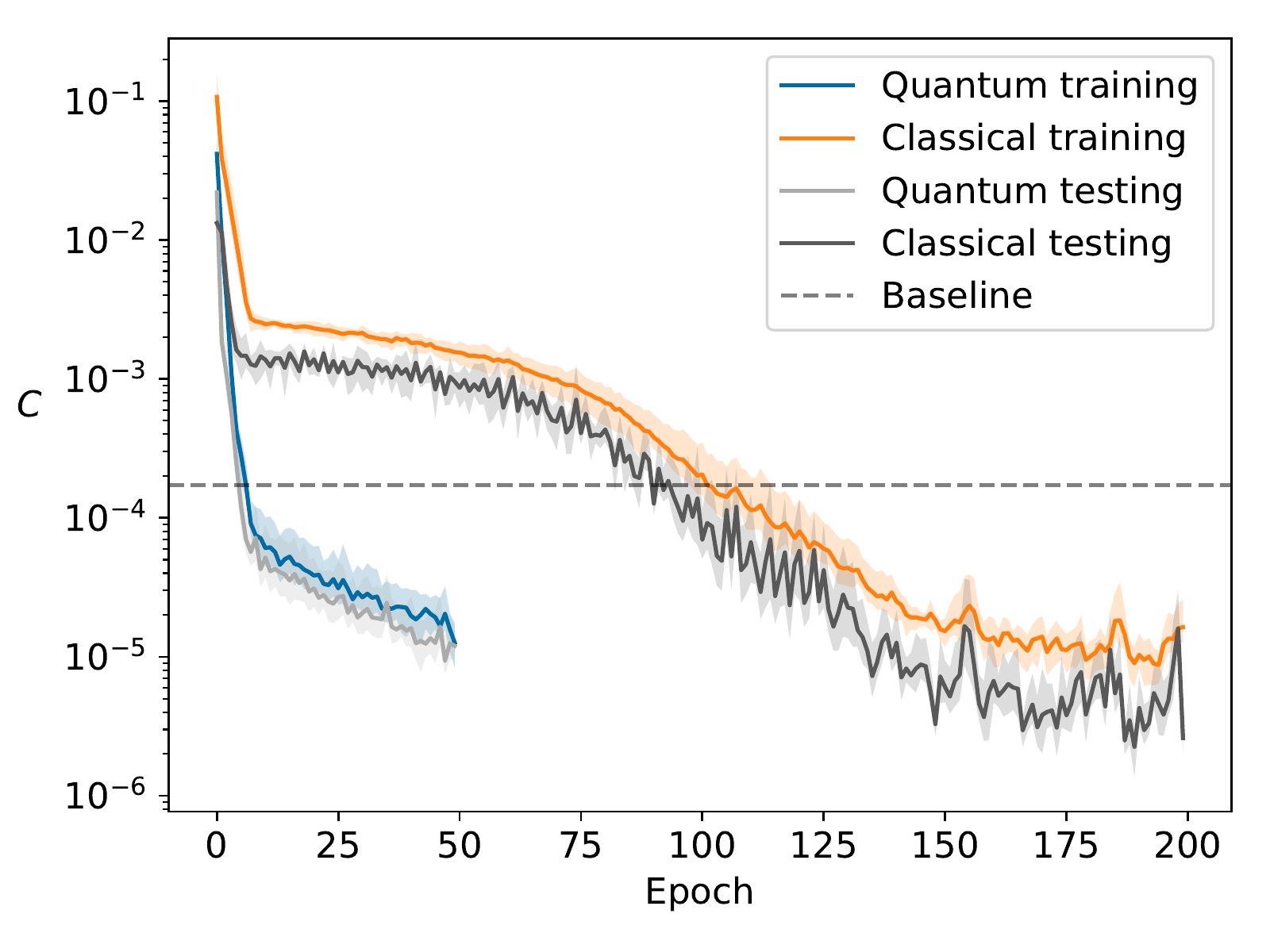}
    \caption{Cost function $C$ (\cref{eq:cost}) computed for CV-QRNN
        (blue line -- training data, light gray -- testing) and LSTM
        (orange line -- training, dark gray -- testing), as a function of
        the number of epochs in the task of predicting the values of the
        Bessel function $J_0(x)$ (Task 1). Shaded regions represent the
        standard deviation and solid lines are the average for 5 runs of
        the simulation.  The CV-QRNN achieves values of $C$ below
        $10^{-4}$ already after 10 epochs and reaches $10^{-5}$ below 50
        epochs. Such values are accessible for the corresponding LSTM
        after 150 epochs. The dashed line indicates the cost function for
        the simplest baseline strategy in which the last input value is
        repeated as the predicted value.} 
    \label{fig:losse_comparison} 
\end{figure}

The results of the first task are depicted in
\cref{fig:losse_comparison}, which shows the cost function $C$
(Eq.~\ref{eq:cost}) as a function of the number of training epochs,
plotted separately for the training and test data sets.  The outputs are
compared for CV-QRNN and LSTM networks, for which we used the same
hyperparameters, such as batch size and learning rate, as well as a
similar number of trainable parameters. The cost function $C$ for the
quantum network reaches the same value after 50 epochs as for the
classical network after 150 epochs. We noticed that in the former case
the cost function drops rapidly in the first few epochs, and for the same
number of epochs it achieves lower values compared to the classical
network.

\begin{figure}[ht]
    \centering
    \begin{subfigure}{0.5\textwidth}
        \includegraphics[width=\textwidth]{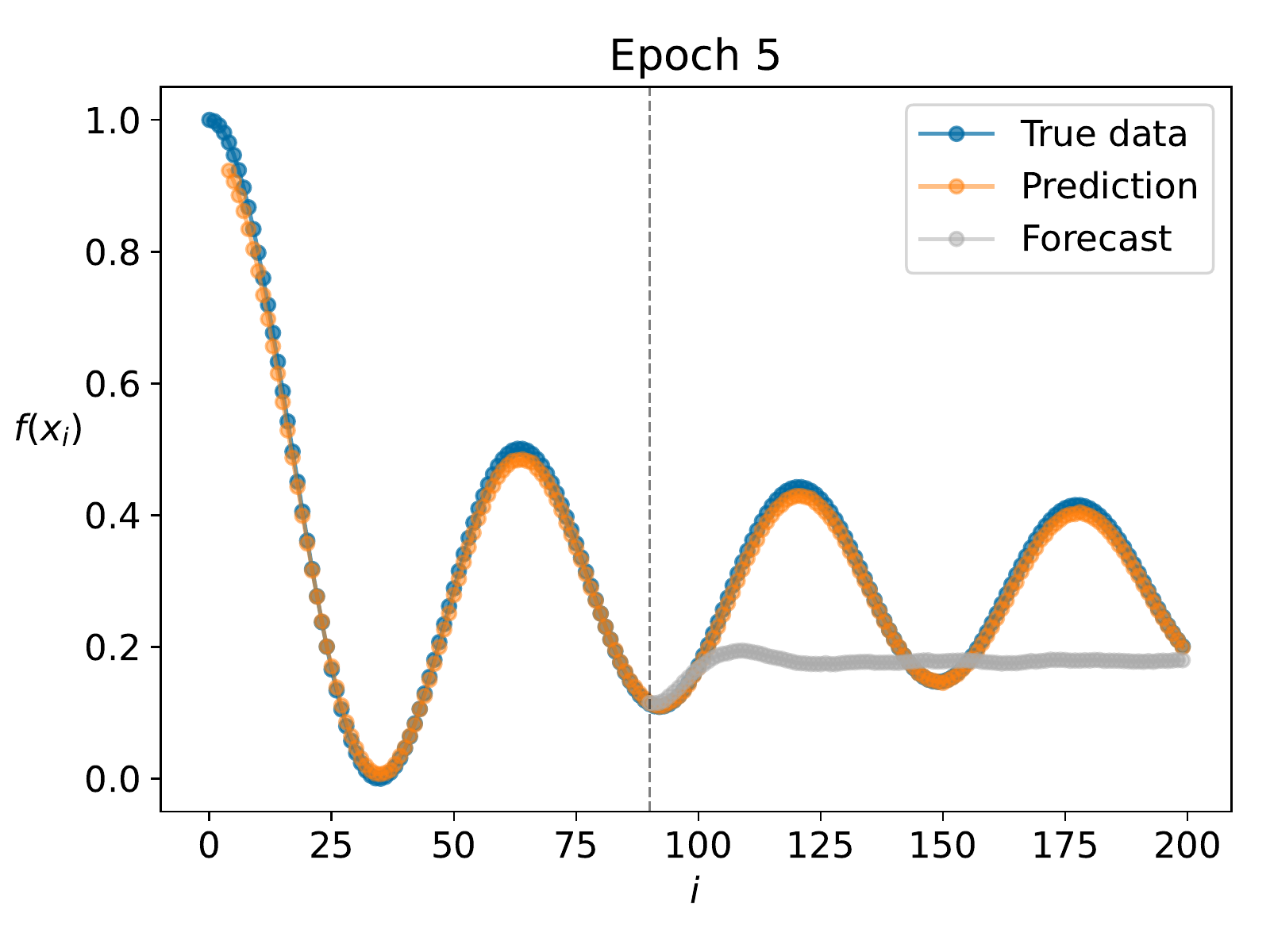}
    \end{subfigure}%
    \begin{subfigure}{0.5\textwidth}
        \includegraphics[width=\textwidth]{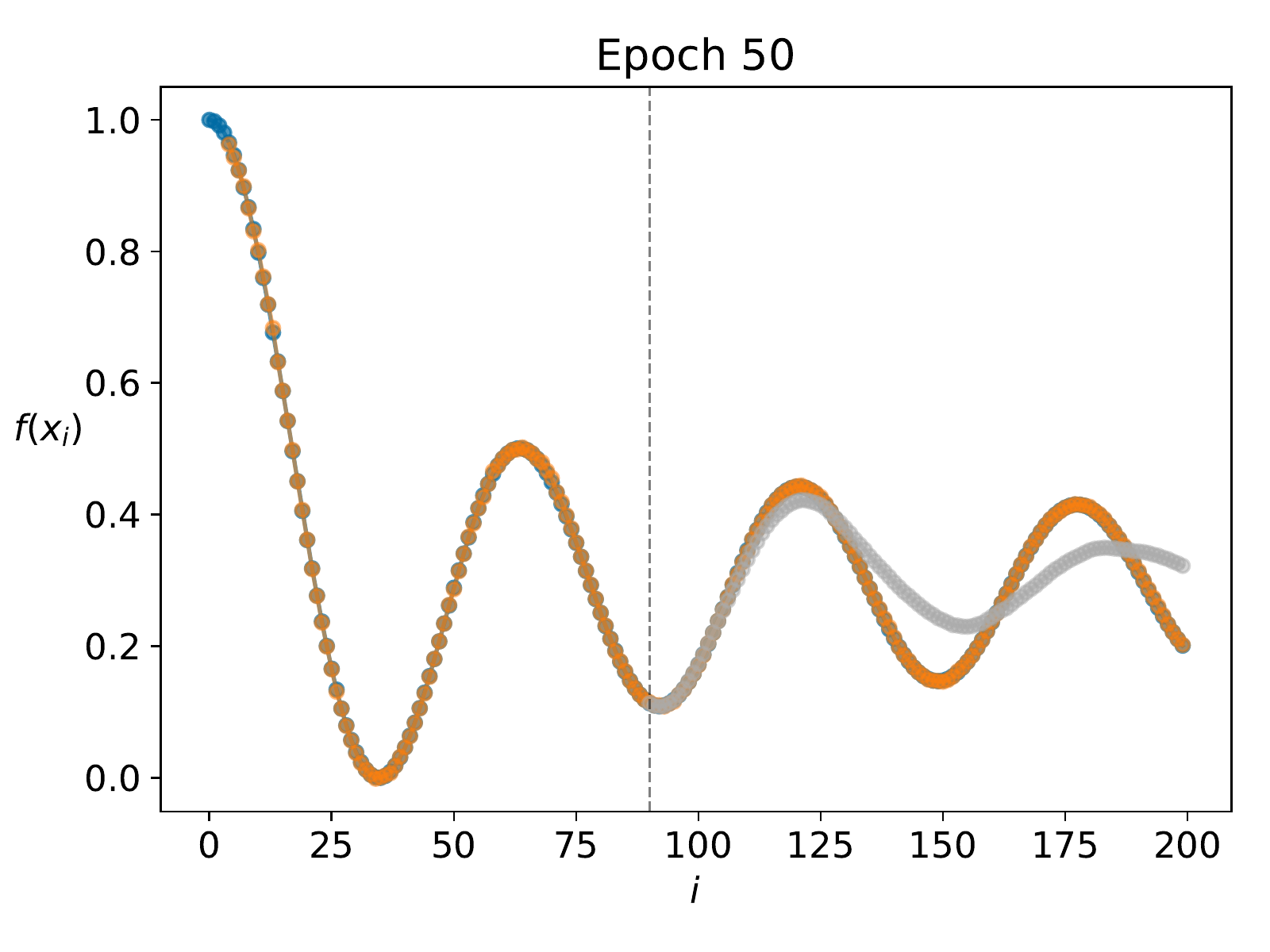}
    \end{subfigure}
    \hfill
    \begin{subfigure}{0.5\textwidth}
        \includegraphics[width=\textwidth]{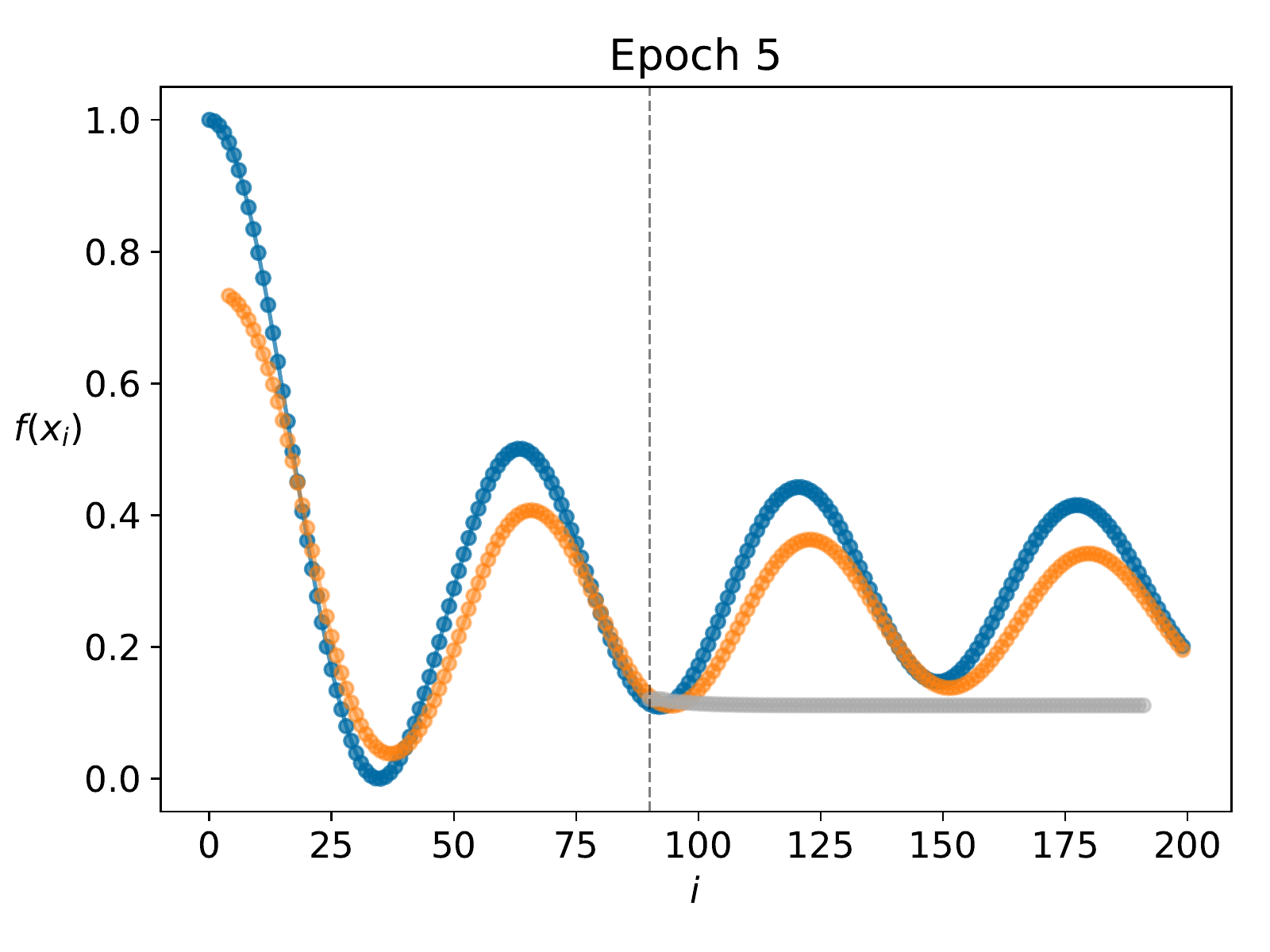}
    \end{subfigure}%
    \begin{subfigure}{0.5\textwidth}
        \includegraphics[width=\textwidth]{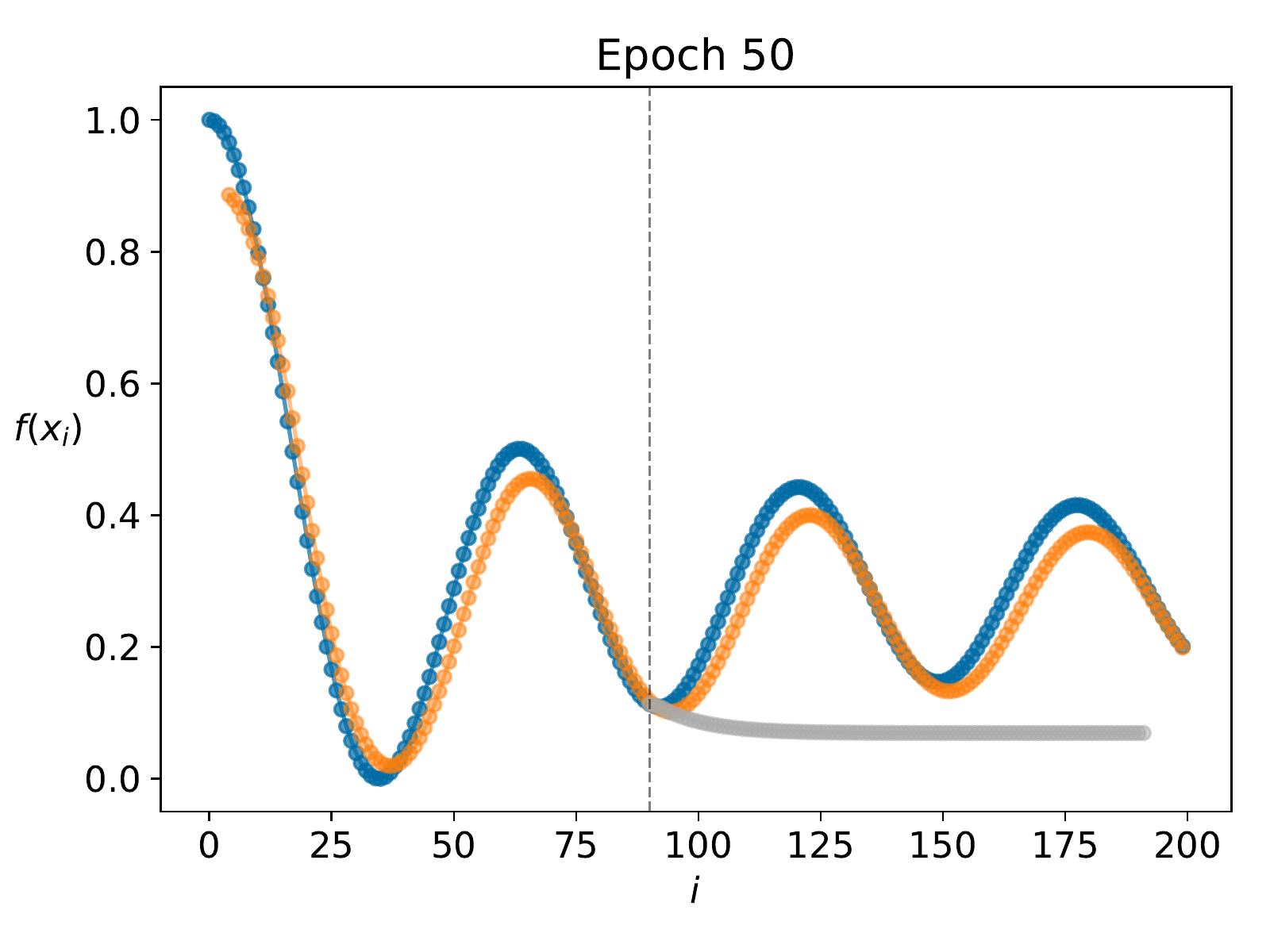}
    \end{subfigure}
    \caption{Progress of training on the data generated with Bessel
        function $J_0(x)$, for CV-QRNN (top row) and LSTM networks
        (bottom row). Blue points represent the reference data, orange
        points are predictions based on $T=4$ previous points, and the
        gray ones -- the forecasted values. Vertical dashed line marks
        the point where the data was split for training (left) and testing
        (right) sequences.}
    \label{fig:bessel_epochs} 
\end{figure}

Prediction and forecasting capabilities of both networks are visualized
in \cref{fig:bessel_epochs}, where output values are compared directly
with the previously generated test sequence. This plot depicts how the
Bessel function is gradually approximated after some number of
computation epochs. It shows that while CV-QRNN copes well with the task
and the prediction is especially well realized, LSTM is much worse in
prediction and fails in forecasting even after 50 epochs of training.

\begin{figure}[t]
  \centering
  \includegraphics[width=0.8\textwidth]{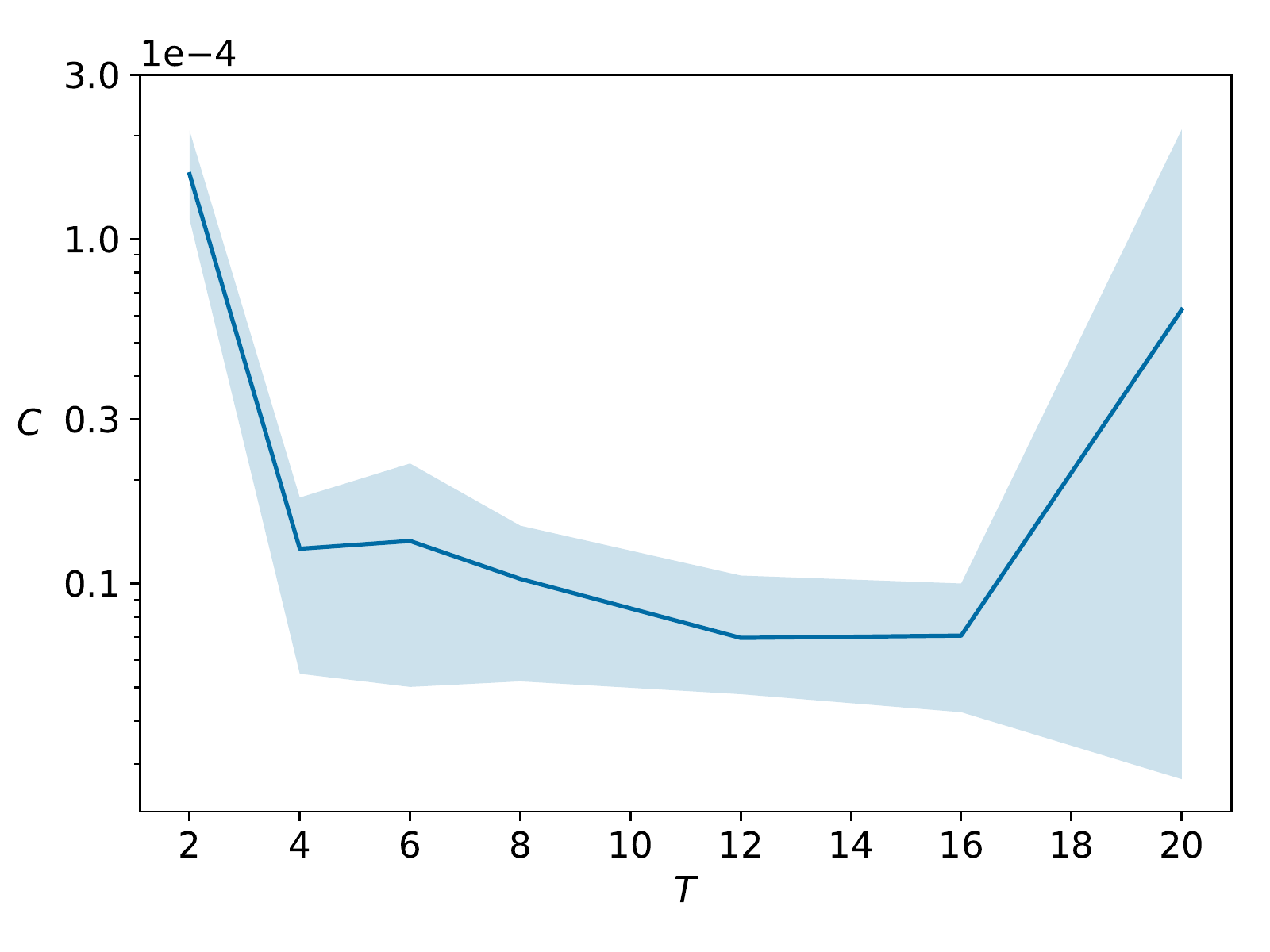}

  \caption{Cost function $C$ (\cref{eq:cost}) after 50 epochs of training
      CV-QRNN for different lengths of input sequence $T$; Mean values
      for 5 separate runs are represented by a line, while the shadow
      depicts the range of achieved values.  Training sequences were
      generated with Bessel function, as described in the text. The
      choice of $T=4$ in our numerical simulations results from the
      observation that for larger lengths the gain is not so large while
      the computing resources and time grows exponentially.}
  \label{fig:bessel_in_len_dependence}
\end{figure}

We also investigated the dependence of CV-QRNN prediction on the input
sequence length $T$, \cref{fig:bessel_in_len_dependence}. For this we
have trained the network with 3 qumodes for $T=2,4,6,8,10,12,16,20$ for
50 epochs and computed the cost function $C$. We observe that the worst
prediction is achieved for $T=2$, which is expected since the recurrent
feature of the network is barely used in this case.  As the $T$
increases, the cost function is decreasing.  This means that the more
information we feed to the network, the better prediction we can obtain.
For $T=20$ we observe large fluctuations in the value of the cost
function, with the best value found being less than for $T=16$ and the
worst -- about the same as for $T=2$. We believe that it is caused by the
cut-off dimension, which is described in \cref{sec:numerical_methods}. In
our experiments, we used $T=4$ which was a compromise between the
computation time and final cost function value.

\begin{figure}[ht]
    \centering
    \includegraphics[width=0.8\textwidth]{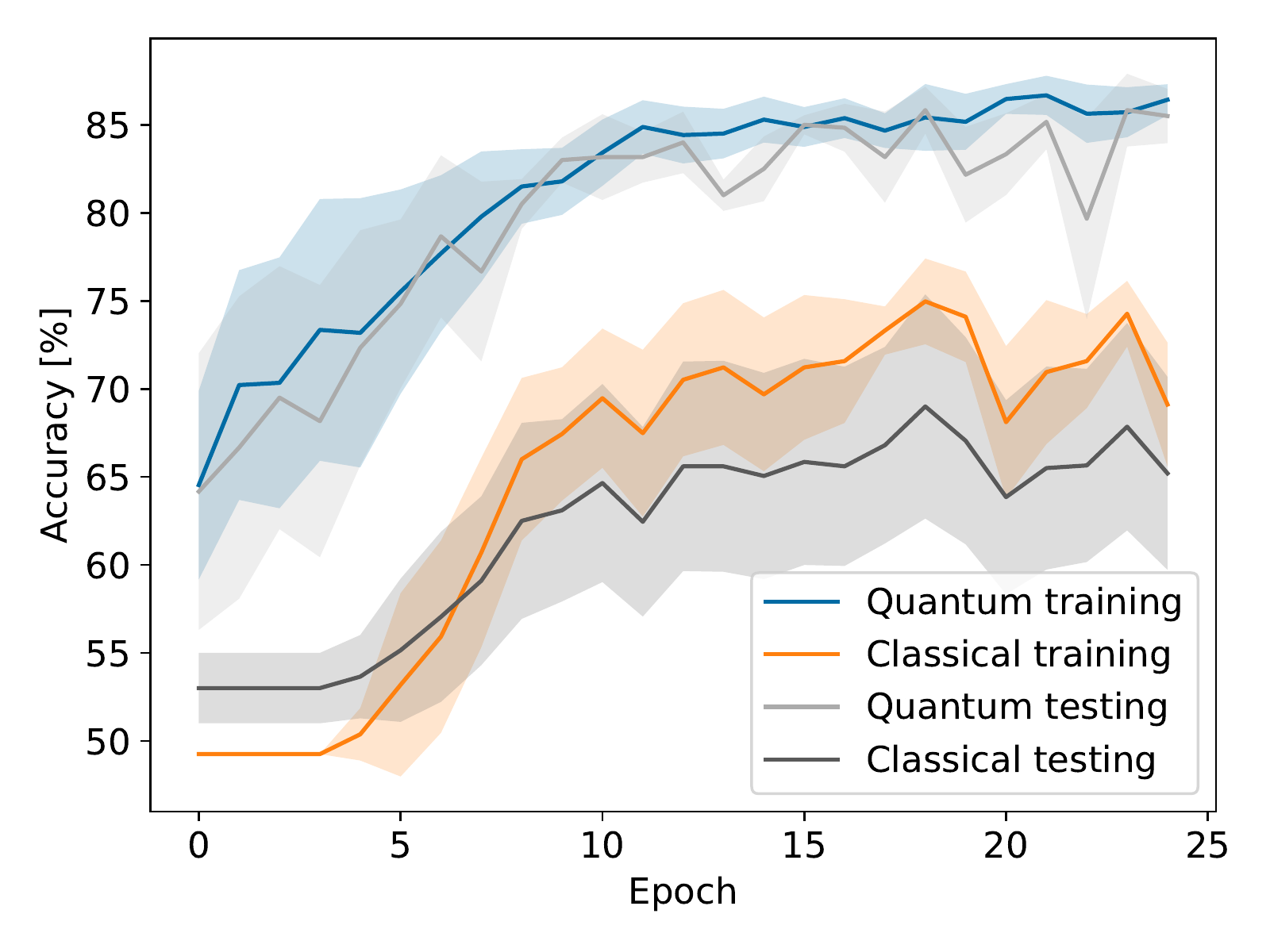}
    \caption{Accuracy (the percentage of properly classified outputs)
        computed for CV-QRNN (blue line -- training data, light gray --
        testing) and LSTM (orange line -- training, dark gray --
        testing), as a function of the number of epochs in the task of
        the classification of the MNIST data set (Task 2). Shaded regions
        represent the standard deviation while solid lines are the
        average for 5 runs of the simulation. The quantum network
        achieves values the accuracy above $80\%$ already after $\sim 10$
        epochs and reaches over $85\%$ in 25 epochs. In the same number
        of epochs the classical network achieves accuracy of $75\%$,
        which does not increase during further training.}
\label{fig:MNIST}
\end{figure}

\textbf{Task 2 -- MNIST image classification.} The second task, which was
tested on the CV-QRNN architecture, was the classification of handwritten
digits from the MNIST data set \cite{lidengMNISTDatabaseHandwritten2012}.
Due to the fact that simulating qumodes and their interactions is
resource-heavy, we have narrowed down the test to the binary
classification problem of digits `3' and `6'. Additionally, we
downsampled the original images from $28 \times 28$ pixels to $7 \times
7$. We used 1000 images, which were divided between training (80\%) and
test (20\%) sets.  The image pixels were sequentially injected into the
network from left to right and from top to bottom, giving the sequence
$\{x_{i}\}_{i=0}^{48}$. The labels were $y \in \{0,1\} $, where $0$
corresponded to digit `3' and $1$ to digit `6'. For the simulations we
used the quantum network with 3 qumodes, with one qumode being an input
qumode and the rest two acting as register modes.  A comparable classical
LSTM network was implemented with the standard machine learning library.
For both quantum and classical networks we have used the binary cross
entropy loss for the calculation of the cost function $C$
(\cref{eq:cost2}).  Additionally, the results were assessed with an
accuracy function, which is defined as the percentage of properly
classified images.

\cref{fig:MNIST} shows the accuracy during the training process for the
MNIST data set. Notably, after $\sim 10$ epochs of training, CV-QRNN
allows one to obtain accuracy above $80\%$. On the other hand, the
classical LSTM architecture with a similar number of parameters arrives
at $75\%$ and this value does not increase with a longer training. LSTM
was trained for 100 epochs, but the accuracy did not change after the
first 25 epochs. Thus, for consistency, we show only the first 25 epochs
for both networks. This result demonstrates a clear advantage of the
quantum network in obtaining similar results after a significantly
smaller number of epochs.


\section{Methods}
\label{sec:numerical_methods}

The quantum network was implemented using the \textit{Strawberry Fields}
package \cite{killoranStrawberryFieldsSoftware2019} that allows the user
to easily simulate CV circuits\footnote{Code is available in the
repository: \url{https://github.com/StobinskaQCAT/CVQRNN}}. It also
provides a backend written in
\textit{TensorFlow}~\cite{abadiTensorFlowLargeScaleMachine2015}, which
makes it possible to use its already implemented functions to optimize
the network parameters. For this purpose, we use the ADAM algorithm,
which is commonly applied to find the optimal parameters of the
network~\cite{zhangImprovedAdamOptimizer2018}. ADAM merges two
techniques: adaptive learning rates and momentum-based optimization. The
initial learning rate was $0.01$ for the time series prediction (Task 1)
and $0.005$ for the classification of MNIST handwritten digits (Task 2).
The data was processed in batches of 7, which allowed us to speed up the
calculation without losing much precision. The hyperparameters were
chosen empirically.

Since the quantum CV computations are done in an infinite-dimensional
Hilbert space, the dimensionality of the system needs to be truncated to
be able to be modeled on a classical computer. The highest accessible
Fock state is called a \textit{cutoff dimension}. In our simulations we
have used the cutoff dimension of 6. Furthermore, we added the
regularization term of the form $L_T = \eta \left( 1 - \text{Tr} \rho
\right)^2 $ to the cost function, where $\text{Tr} \rho$ is the trace of
the state after the last layer has been processed, and $\eta$ is a weight
empirically chosen to be 10
\cite{killoranContinuousvariableQuantumNeural2019}.

The implementation of classical LSTM has been realized using the
\textit{TensorFlow} package \cite{abadiTensorFlowLargeScaleMachine2015}.
We use the layer \texttt{tf.keras.layers.LSTM}, which takes as a
parameter the dimensionality of the hidden state. We set this parameter
to match the number of trainable parameters in CV-QRNN, to make both
implementations comparable. The remaining arguments of the LSTM
implementation were left at default values.

The calculations were performed with two hardware platforms. The time
series predcition task (Task 1) was realized on a laptop with CPU Intel
Core i5-10210U (8 cores) running at 4.2GHz, and 16 GB of RAM. The
calculations took between 1 to 24 hours for CV-QRNN training over 50
epochs, depending on the data input length. The training for the MNIST
data classification (Task 2) was realized with a cluster with CPU Intel
Xeon E5-2640 v4 processors, 120 GB of RAM and Titan V GPUs equipped with
128 GB of memory. It took approximately two days for 25 epochs and 1000
images of 49 pixels each.


\section{Discussion} \label{sec:results}

We performed extensive numerical simulations of CV-QRNN with a CV quantum
simulator software and compared its performance to the state-of-the-art
implementation of classical LSTM. Our simulations showed that CV-QRNN
possesses features that make it highly advantageous in time series
processing compared to the classical network. The quantum network arrived
at its optimal parameter values (cost function below $10^{-5}$) within 50
epochs, while a comparable classical network achieved the same goal after
150 epochs, and therefore the speed gain achieved 300\%. Similar results
were obtained for other sets, presented in
Appendix~\ref{sec:additional_datasets}.

Faster RNN training is a hot topic currently investigated by AI
researchers~\cite{garcia-martinEstimationEnergyConsumption2019}, who
notice that it becomes a more important goal than achieving high
accuracy. High requirements for computing power and energy consumption in
large machine learning models constitute serious roadblocks for their
deployment. They directly translate into large operational costs, but
also into an environmental footprint, and therefore, they must be
resolved. Therefore, the computation speedup merged with lower
environmental influence are unbeatable advantages of quantum platforms
which directly address the limitations faced by classical solutions.

Our work opens possible prospects for future research in the development
of quantum RNNs. It also underlines the importance of the CV quantum
computation model.  The quantum platform we chose makes our solution
highly compelling, because the CV architecture we propose can be
implemented with existing off-the-shelf quantum photonic hardware, which
operates at room temperature. To develop such a platform, one needs
lasers, which produce coherent sources of light, and basic elements
(squeezers, phase shifters, beam splitters), which are already routinely
implemented in photonic chips and are characterized by very low losses.
Homodyne detection achieves very high efficiency and is implemented with
photodiodes and electronics.  To obtain suitable nonlinearity, required
for the activation function, we used the tensor product structure of the
quantum circuit together with the measurement, which freed us from using
additional nonlinear elements. One of the unbeatable advantages of this
platform is true quantum operation, without the need of performing
measurements of the reference qubits or qumodes inside each algorithm
iteration to implement internal rules of the hidden layer. However,
measurement of selected qumodes and use of this output for a subsequent
iteration is perfectly doable. A similar approach was already
demonstrated in the Coherent Ising Machine and is planned for its quantum
successor~\cite{inagakiCoherentIsingMachine2016,
yamamuraQuantumModelCoherent2017, honjo100000spinCoherent2021}. There, a
very long optical fiber loop acted as a delay line to synchronize
electronic and photonic paths of the circuit.

A natural next step for our project would be to repeat the computations
with real quantum hardware instead of a simulator. This is the problem
faced by many scientific papers in the domain of quantum machine
learning, as they usually do not rely on one of a few available hardware
configurations. For example, previously studied quantum RNN architectures
such as QLSTM or QGRU relied on unphysical operations such as copying of
a quantum state, which could not be achieved with real quantum hardware.

There are also several open questions that would be worth answering in
future work. One of them is the framework in which a comparison between
classical and quantum networks would be possible in a fair way. In our
work, we used the criterion of the same number of parameters, however,
there are approaches that focus on provable advantages of a quantum
network \cite{gyurikEstablishingLearningSeparations2022,
huangQuantumAdvantageLearning2022}.  Moreover, our simulations, due to
the availability of limited computational resources and exponential
scaling of requirements, were performed only for a small number of
qumodes. Therefore, in future research, we would like to verify if a
similar qunatum advantage is still present for a larger network. Lastly,
it would be particularly interesting to study CV-QRNN performance with
real-world data such as hurricane intensity
\cite{giffard-roisin2018ClimateInformatics2018}, where a clear data
pattern is not obvious. 


\section{Acknowledgments}

M.S. and M.St. were supported by the National Science Centre ``Sonata
Bis'' project No. 2019/34/E/ST2/00273. A.B. and M.St. were supported by
the European Union’s Horizon 2020 research and innovation programme under
the Marie Skłodowska-Curie project ``AppQInfo'' No. 956071. In addition,
M.St. was supported by the QuantERA II Programme that has received
funding from the European Union’s Horizon 2020 research and innovation
programme under Grant Agreement No 101017733, project ``PhoMemtor'' No.
2021/03/Y/ST2/00177. We thank T. McDermott for early discussions on
literature findings.


\section{Declarations}

\subsection{Competing interests}

The authors declare no competing interests.

\subsection{Authors' contributions}

M.Si. has wrote the computer program for simulation, conducted numerical calculations and
prepared all figures. B.LS. and M.St. helped with the literature review, created the idea
of CV-RNN and theoretically support it. A.B., M.Si., M.St. and B.LS. wrote the manuscript.

\subsection{Availability of data and materials}

The code used for the simulations and to create the plots is available on 
\url{https://github.com/StobinskaQCAT/CVQRNN}.


\bibliography{biblio}

\clearpage


\appendix

\section{Noise influence}
\label{sec:noise}

\begin{figure}[ht]
  \centering
  \begin{subfigure}{0.7\textwidth}
    \begin{center}
      \includegraphics[width=\textwidth]{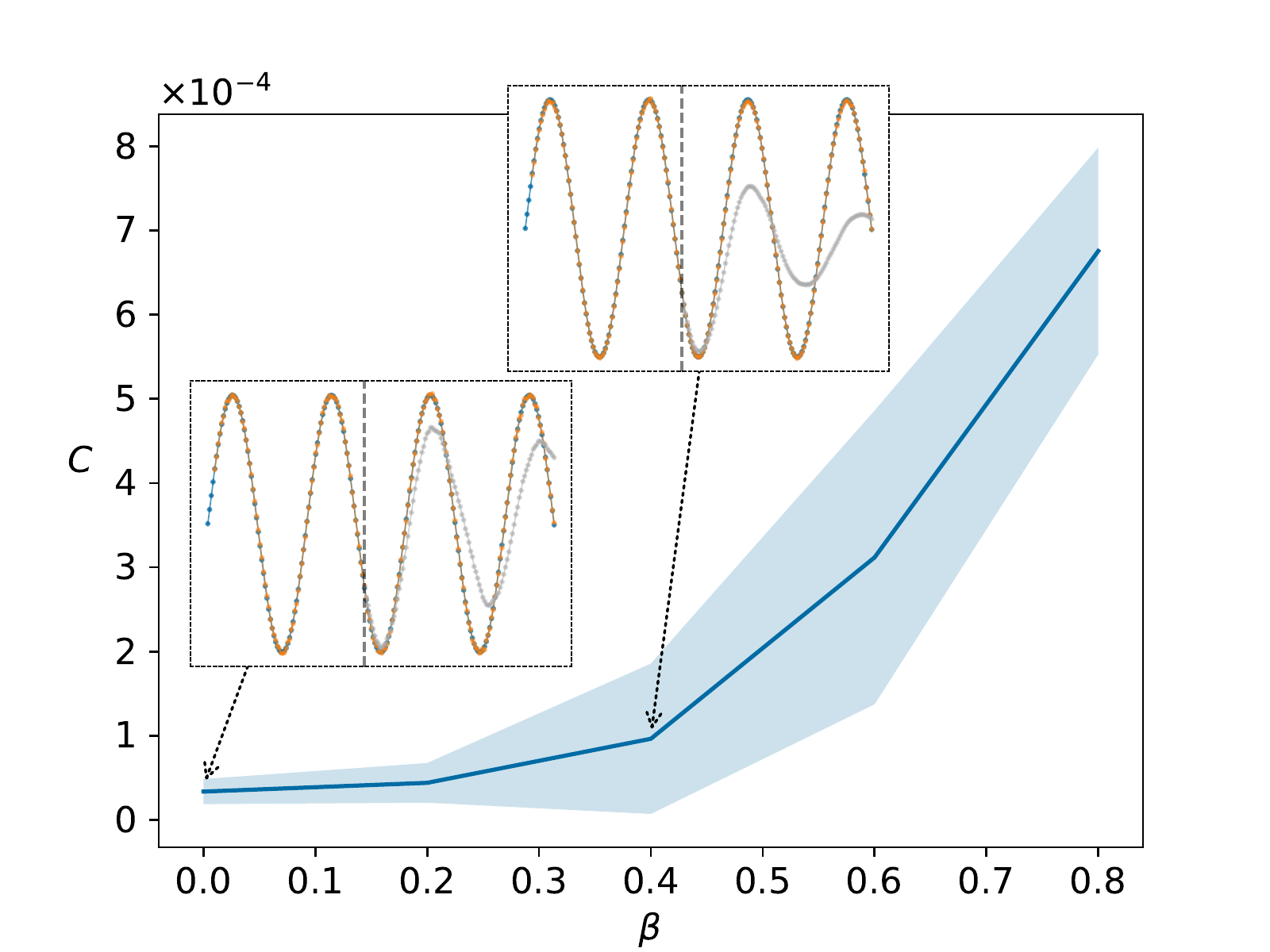}
    \end{center}
    \caption{}
    \label{fig:channel_noise}
  \end{subfigure}
  \begin{subfigure}{0.7\textwidth}
    \begin{center}
      \includegraphics[width=\textwidth]{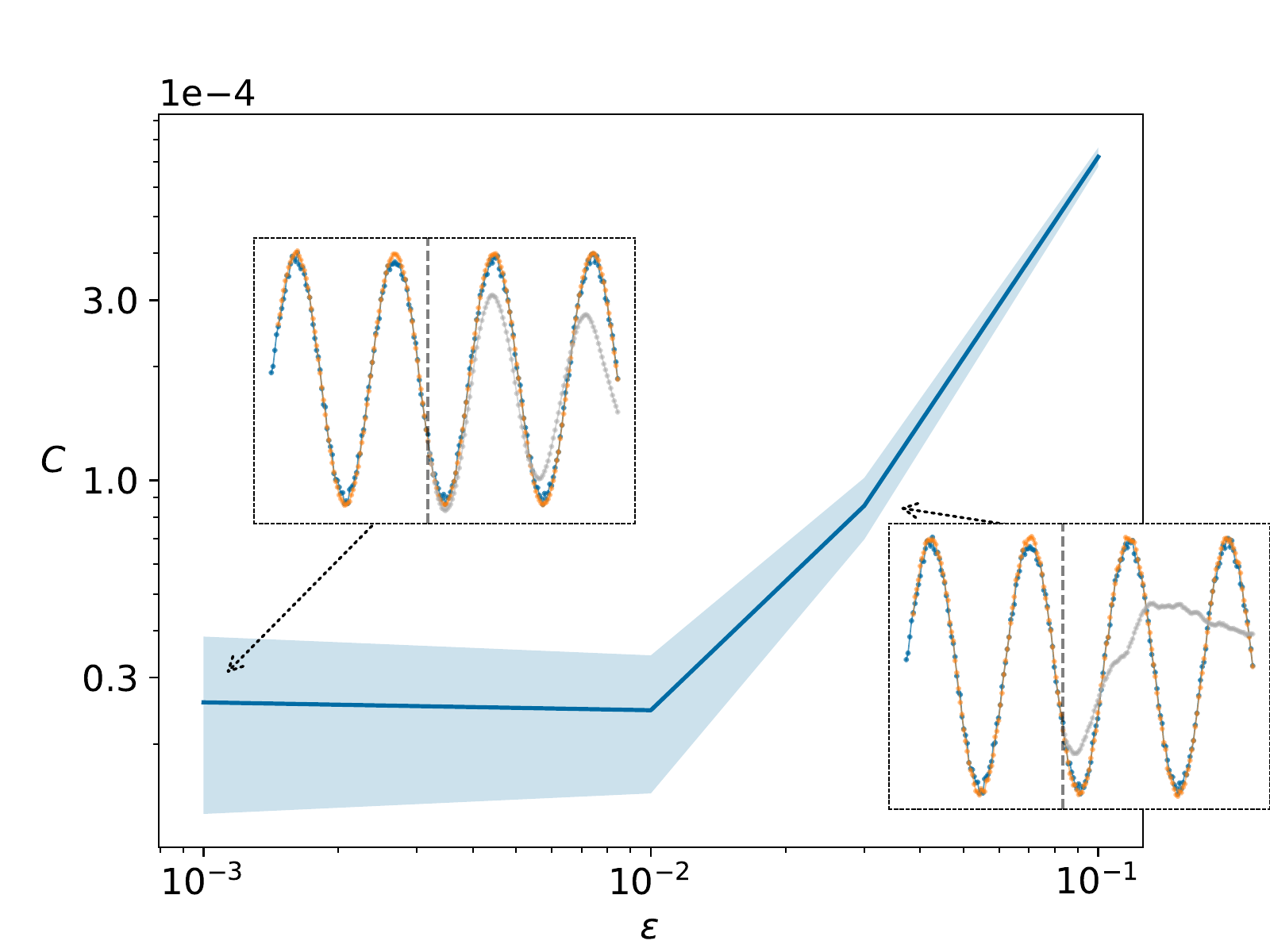}
    \end{center}
    \caption{}
    \label{fig:data_noise}
  \end{subfigure}
  \caption{The influence of the noise on the cost function $C$
    (\cref{eq:cost}): (a) for noise in the channel parametrized by
    $\beta$ and (b) for noise in the data, parametrized by $\varepsilon$.
    Both parameters are described in \cref{sec:noise}.  Shaded region
    shows the standard deviation, while solid lines depicts a mean of 5
    runs of the simulation. In the small boxes, the prediction of the
    network for the parameter shown by dashed arrow are presented.  For
    clarity we have omitted legends, but the colors are the same as in
  \cref{fig:bessel_epochs}}
\end{figure}

We analyzed the influence of noise on the predictions returned by
CV-QRNN in Task 1. Two types of noise were investigated.  The first was
caused by losses in the channels and was estimated in the following
way: let $\hat{a}$ be a bosonic mode and $\beta \in [0,1]$ be the loss
parameter; then the lossy channel is described by
\begin{equation} 
  \hat{a}' = \text{Tr}_{\hat{b}} \left( \sqrt{1-\beta}~\hat{a} +
  \sqrt{\beta}~\hat{b} \right),
\end{equation}
where $1-\beta$ is \textit{energy transitivity}. For $\beta=0$ we obtain
the original mode $\hat{a}$, and for $\beta=1$ we lose all the
information. The dependence of final cost function on parameter $\beta$
is shown in \cref{fig:channel_noise}.

The other type of noise is located in the data itself. To model it, we
added uniformly distributed random values to the time series,
$\text{Uniform}(-\varepsilon, \varepsilon)$, where $\varepsilon$ is a
parameter. The dependence of final cost function on the parameter
$\varepsilon$ is presented in \cref{fig:data_noise}.

Importantly, we did not observe significant change in the prediction
for a network with a channel loss up to 0.2. The cost function for
$\beta = 0.4$ is twice as big as for no noise at all. The network still
performs well, even the forecasting of multiple data points (cf.\
\cref{fig:channel_noise}). In the case of the noisy data we observed no
influence up to $\varepsilon = 0.01$. With $\varepsilon = 0.03$ we
obtained the value of the loss function, which is almost an order of
magnitude larger than with no loss at all. \cref{fig:data_noise} also
shows that for $\varepsilon > 0.03$ the network loses its ability to
forecast.

\section{Additional data sets}
\label{sec:additional_datasets}

\begin{figure}[ht]
  \centering
  \begin{subfigure}{0.49\textwidth}
    \begin{center}
      \includegraphics[width=1\textwidth]{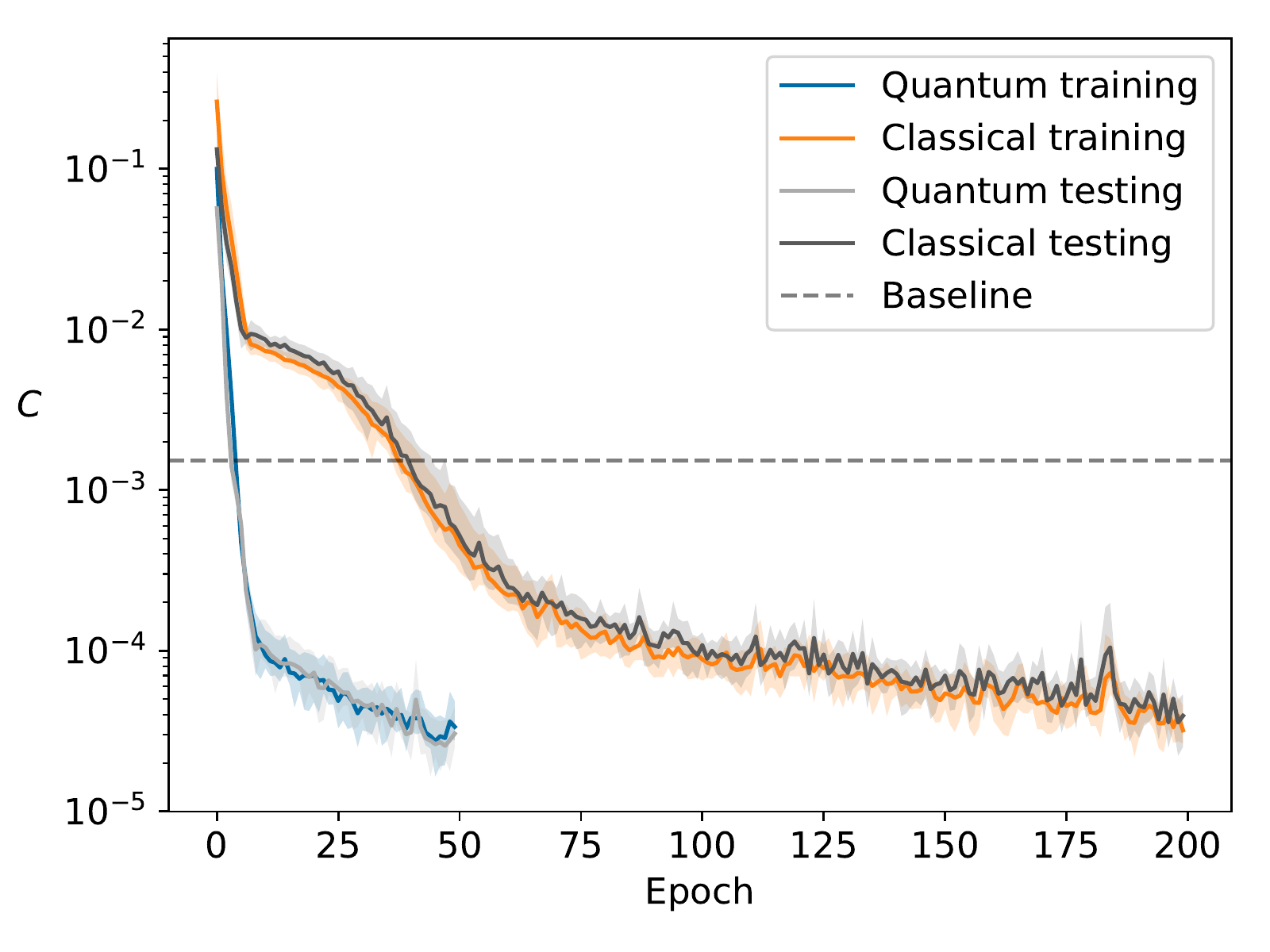}
    \end{center}
    \caption{}
    \label{fig:sin_loss}
  \end{subfigure}
  \begin{subfigure}{0.49\textwidth}
    \begin{center}
      \includegraphics[width=1\textwidth]{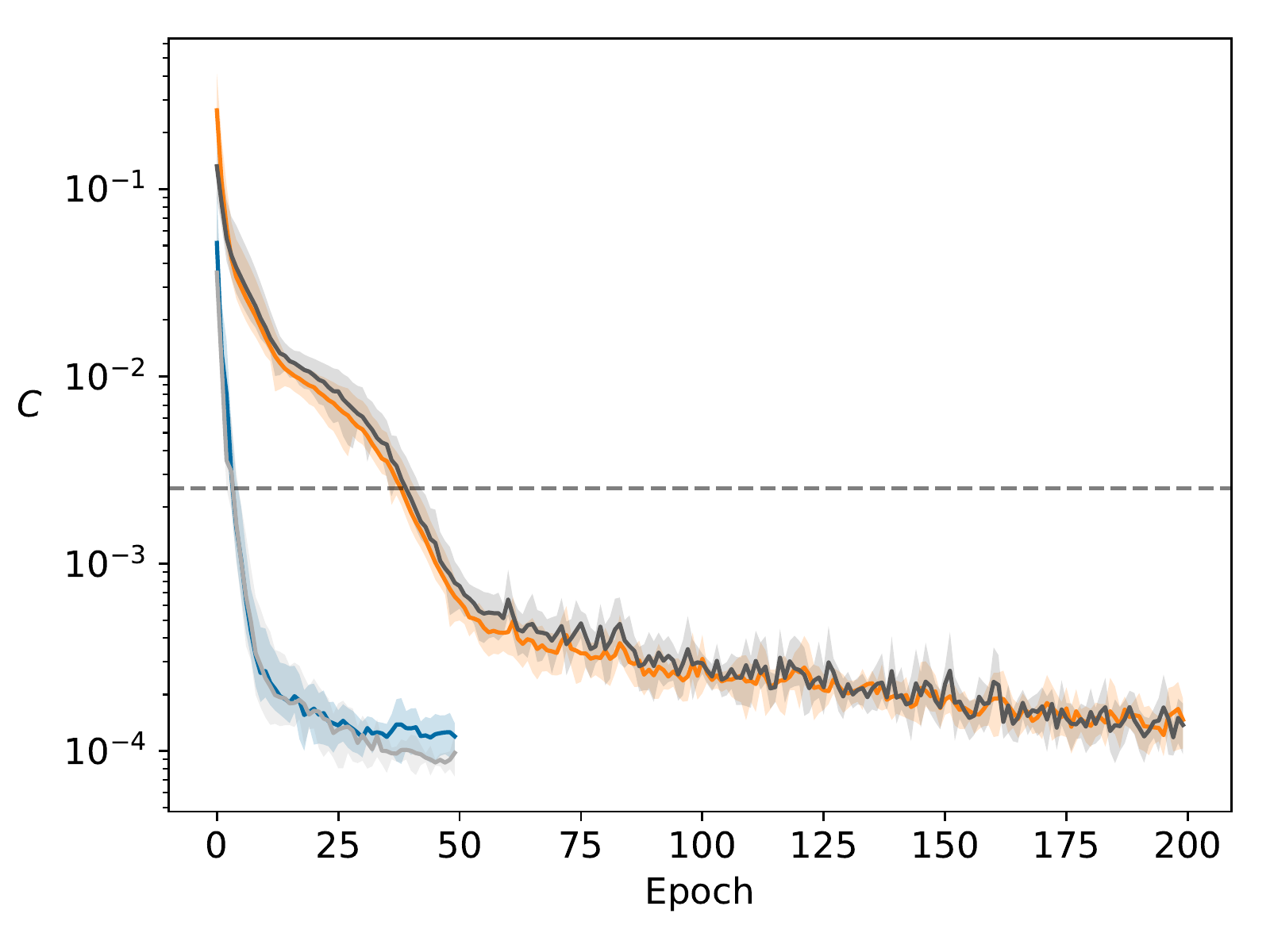}
    \end{center}
    \caption{}
    \label{fig:sin_loss_2}
  \end{subfigure}
  \begin{subfigure}{0.49\textwidth}
    \begin{center}
      \includegraphics[width=1\textwidth]{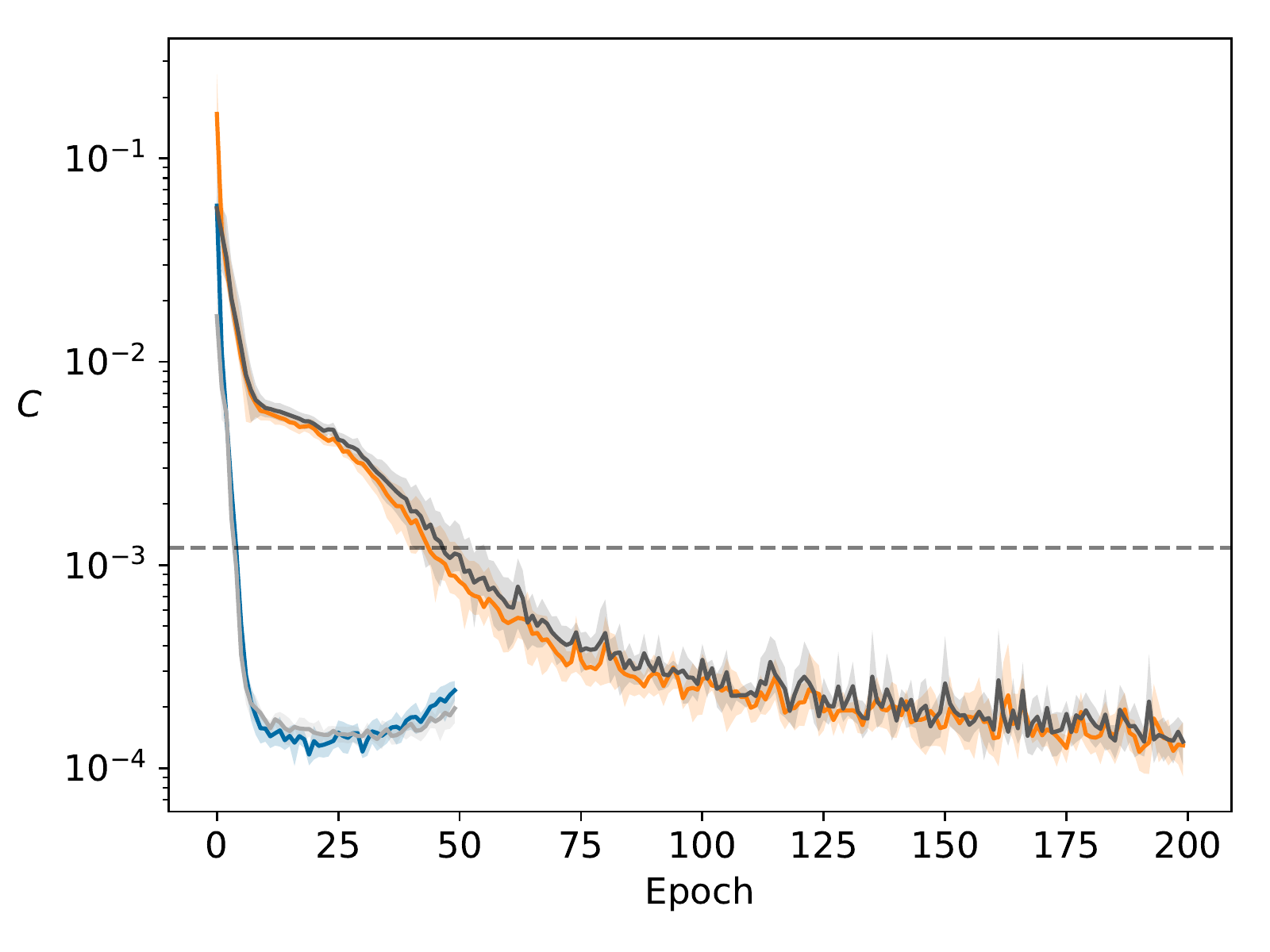}
    \end{center}
    \caption{}
    \label{fig:triangle_loss}
  \end{subfigure}
  \begin{subfigure}{0.49\textwidth}
    \begin{center}
      \includegraphics[width=1\textwidth]{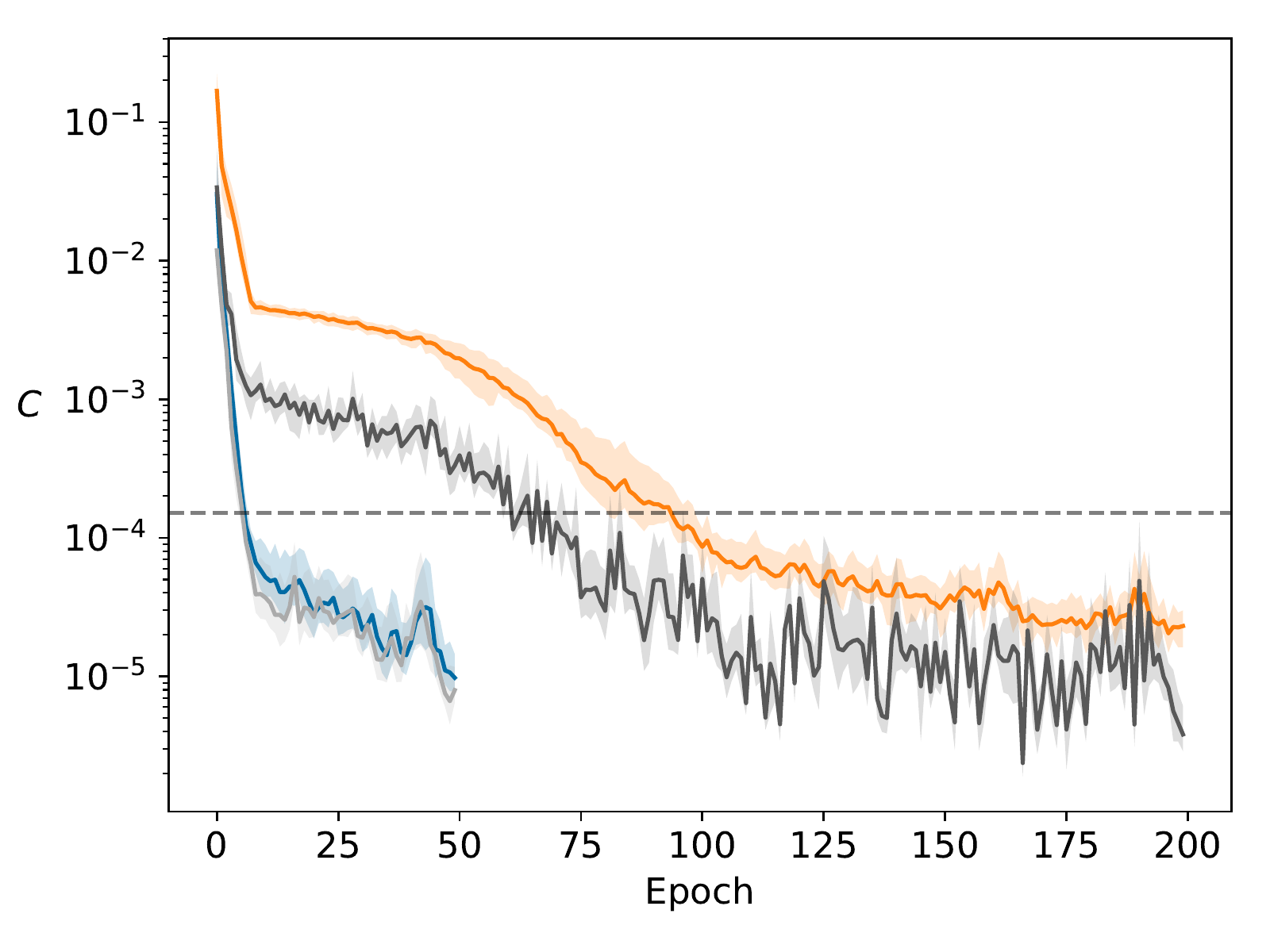}
    \end{center}
    \caption{}
    \label{fig:cos_loss}
  \end{subfigure}
  \caption{Cost functions $C$ (\cref{eq:cost}) during the training of the
    network for different data sets: (a) sine wave, (b) composition of 2
    sine waves with different periods, (c) triangle wave, (d)
    exponentially damped cosine wave.}
  \label{fig:loss_rest}
\end{figure}

\begin{figure}[ht]
  \centering
  \begin{subfigure}{0.5\textwidth}
    \includegraphics[width=\textwidth]{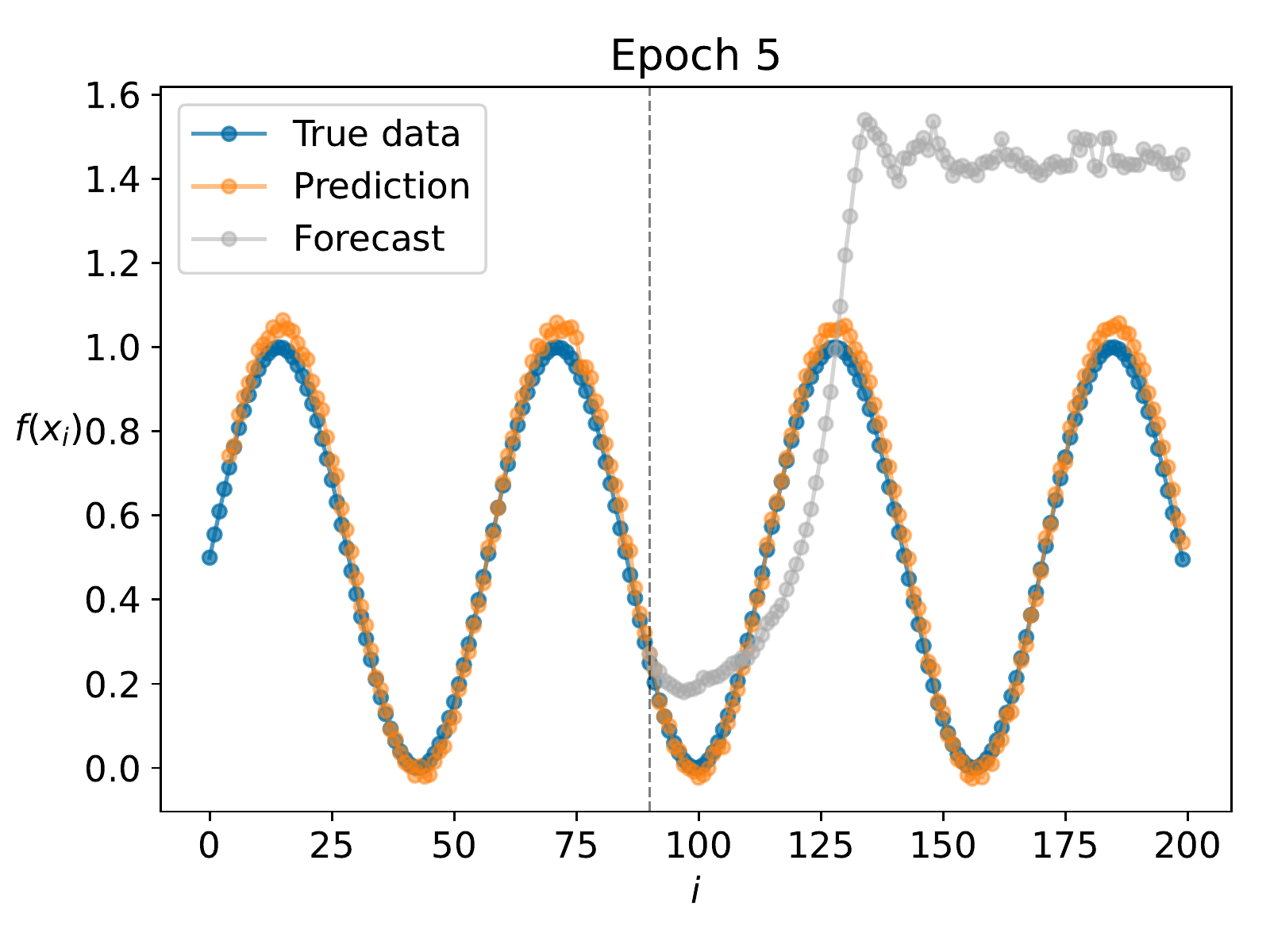}
  \end{subfigure}%
  \begin{subfigure}{0.5\textwidth}
    \includegraphics[width=\textwidth]{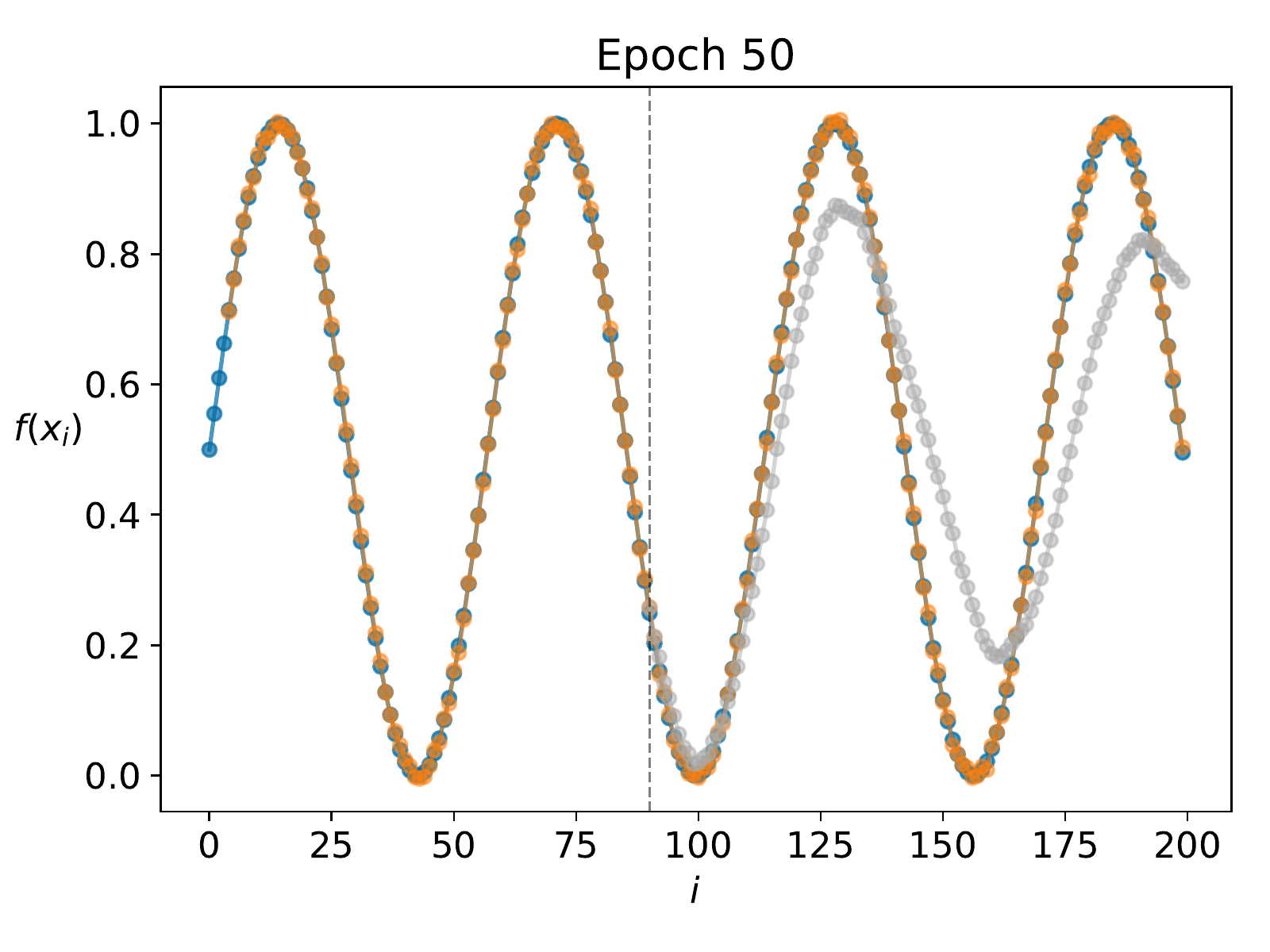}
  \end{subfigure}
  \hfill
  \begin{subfigure}{0.5\textwidth}
    \includegraphics[width=\textwidth]{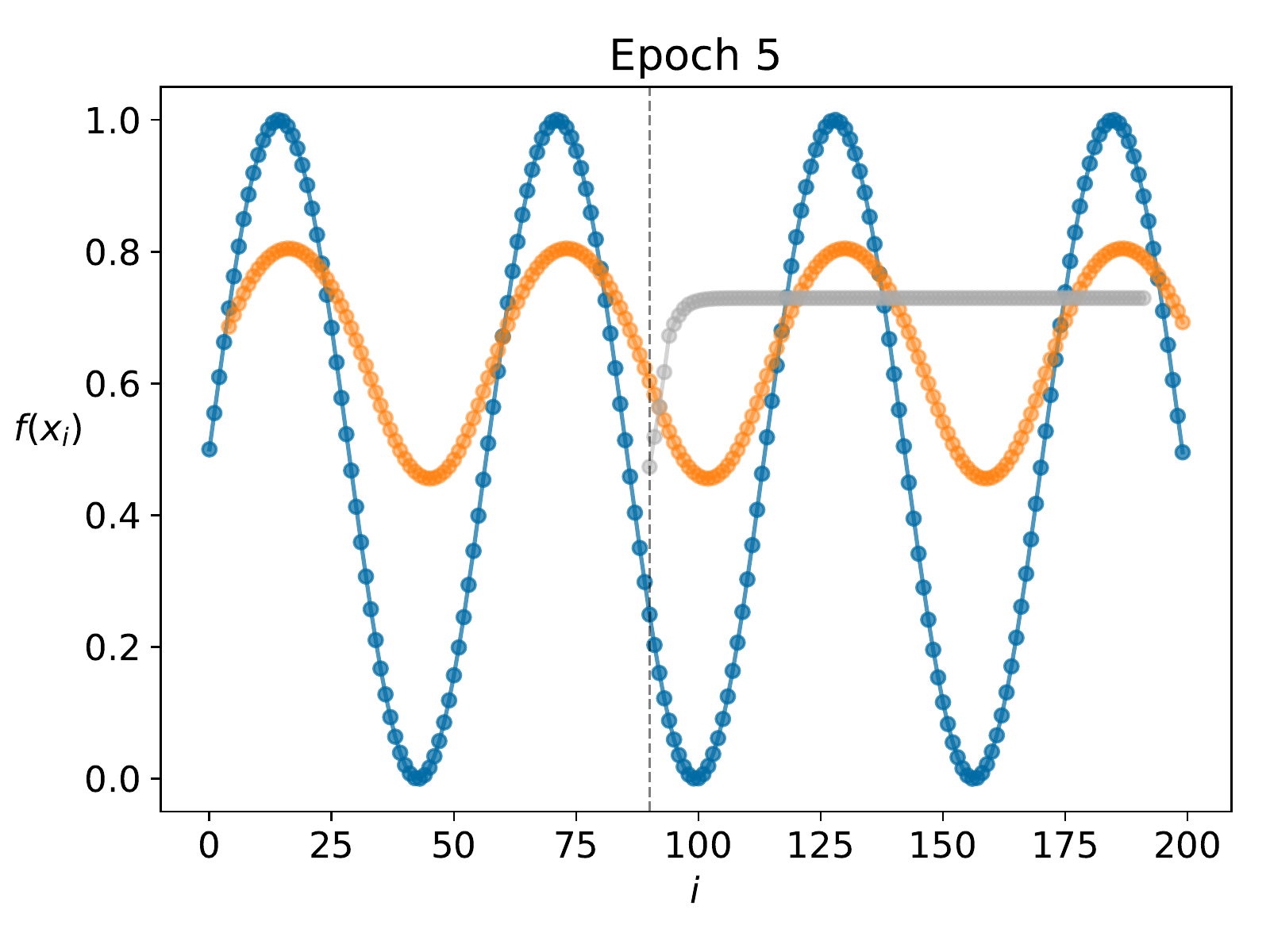}
  \end{subfigure}%
  \begin{subfigure}{0.5\textwidth}
    \includegraphics[width=\textwidth]{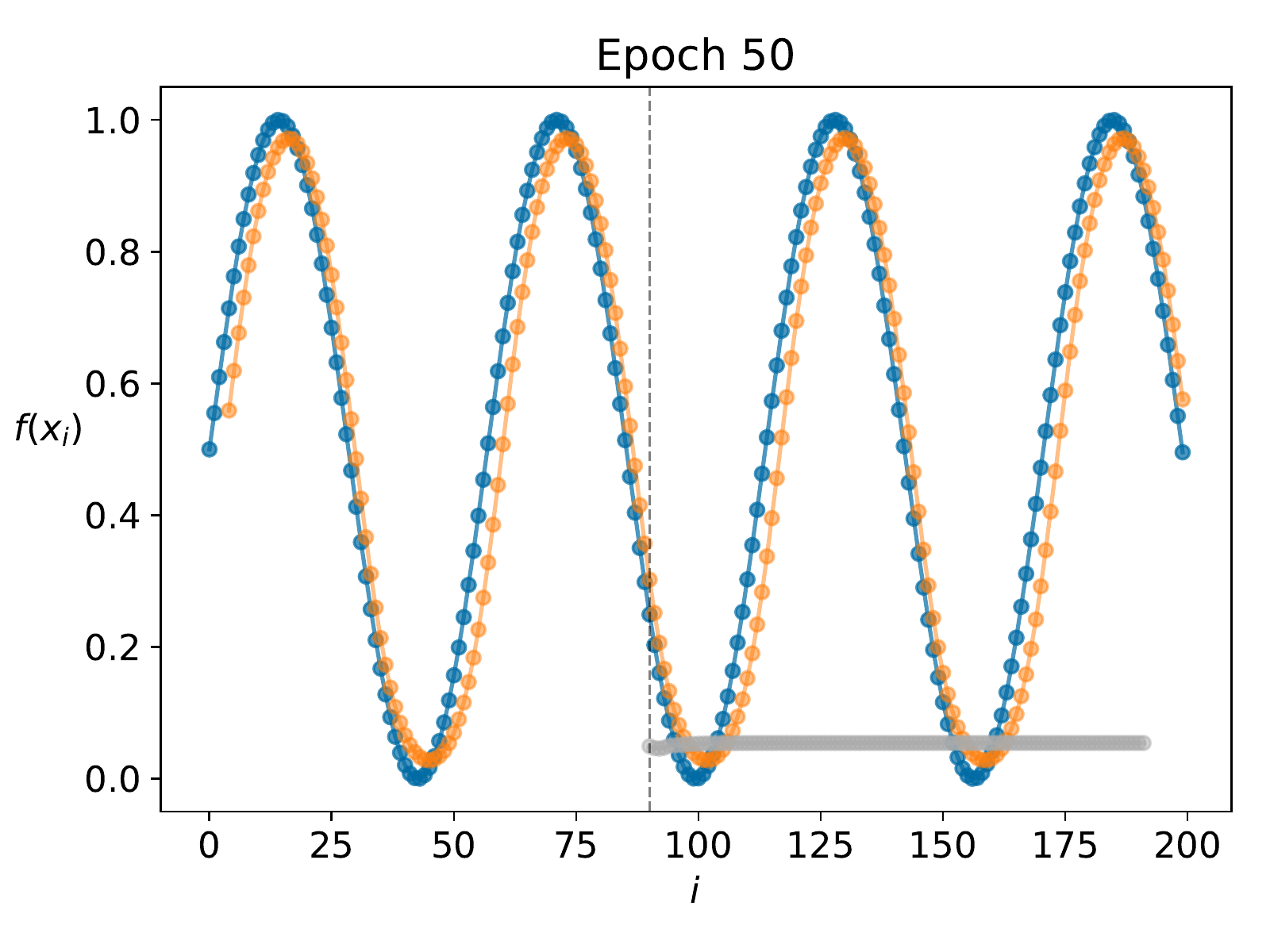}
  \end{subfigure}
  \caption{Progress of training on the data generated with sine
    function $\sin(x)$, for CV-QRNN (top row) and LSTM networks
    (bottom row). Blue points represent the reference data, orange
    points are predictions based on $T=4$ previous points, and the
    gray ones -- the forecasted values. Vertical dashed line marks
    the point where the data was split for training (left) and testing
  (right) sequences.}
  \label{fig:sine_wave_epochs} 
\end{figure}

\begin{figure}[ht]
  \centering
  \begin{subfigure}{0.5\textwidth}
    \includegraphics[width=\textwidth]{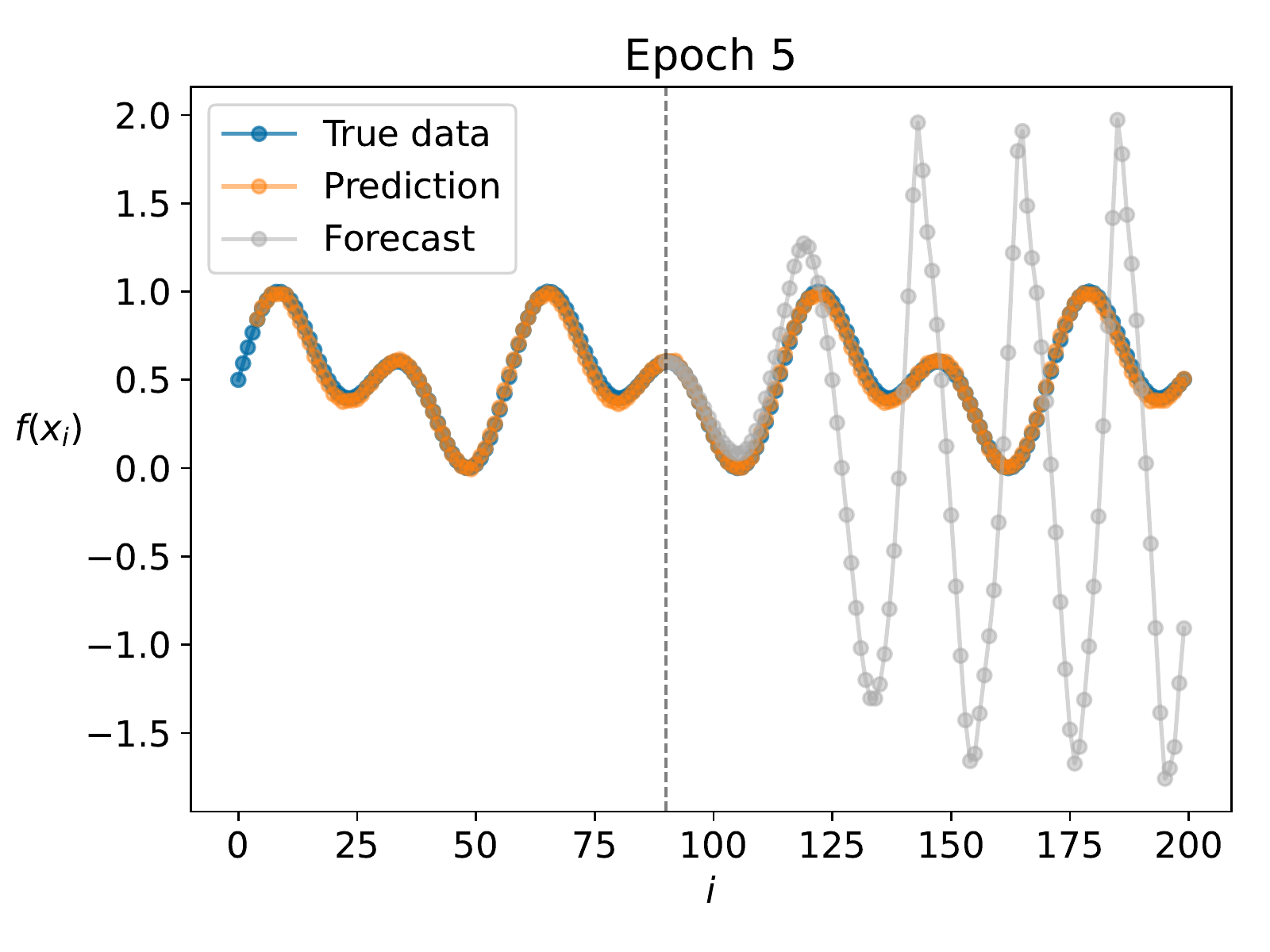}
  \end{subfigure}%
  \begin{subfigure}{0.5\textwidth}
    \includegraphics[width=\textwidth]{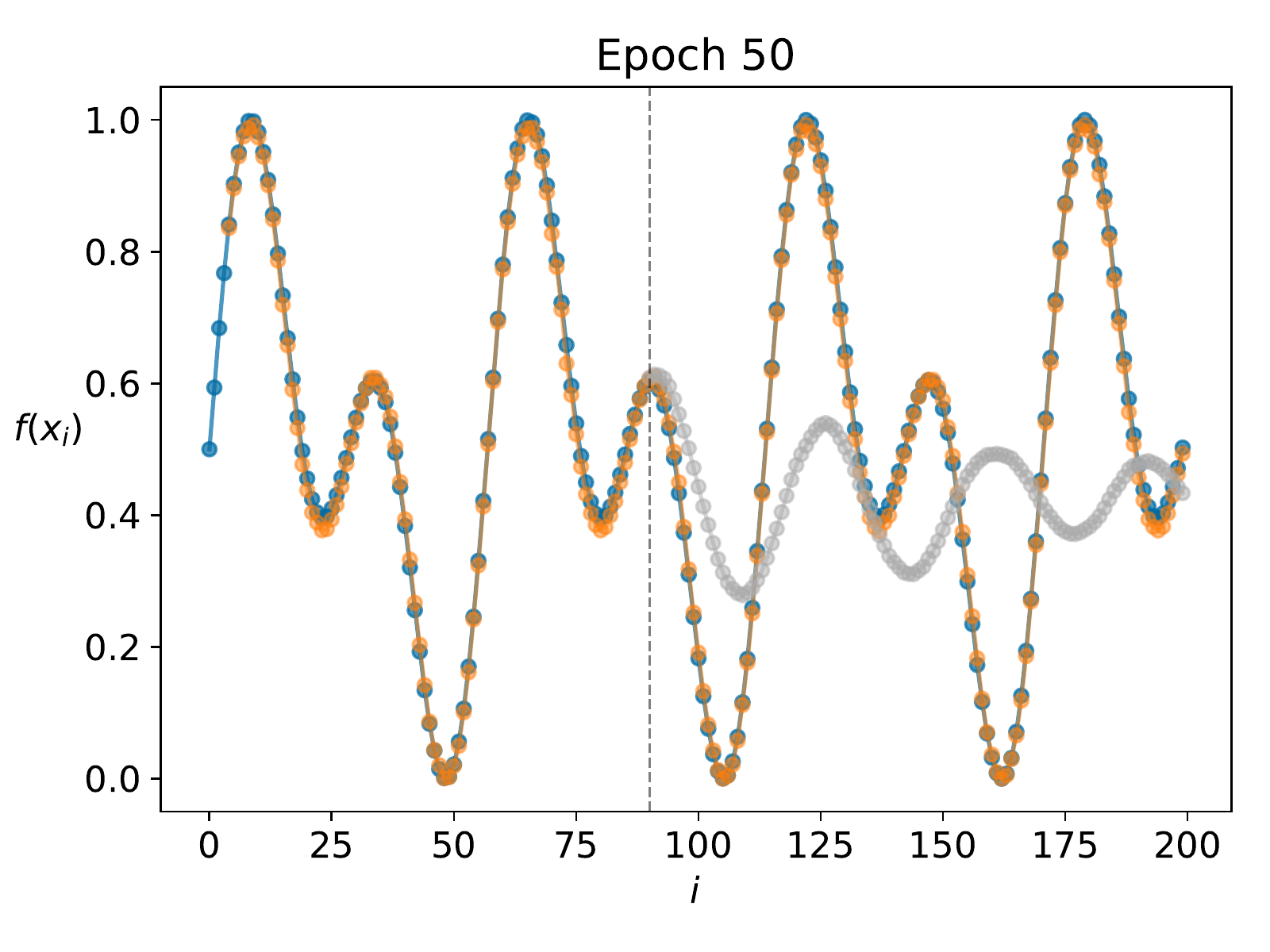}
  \end{subfigure}
  \hfill
  \begin{subfigure}{0.5\textwidth}
    \includegraphics[width=\textwidth]{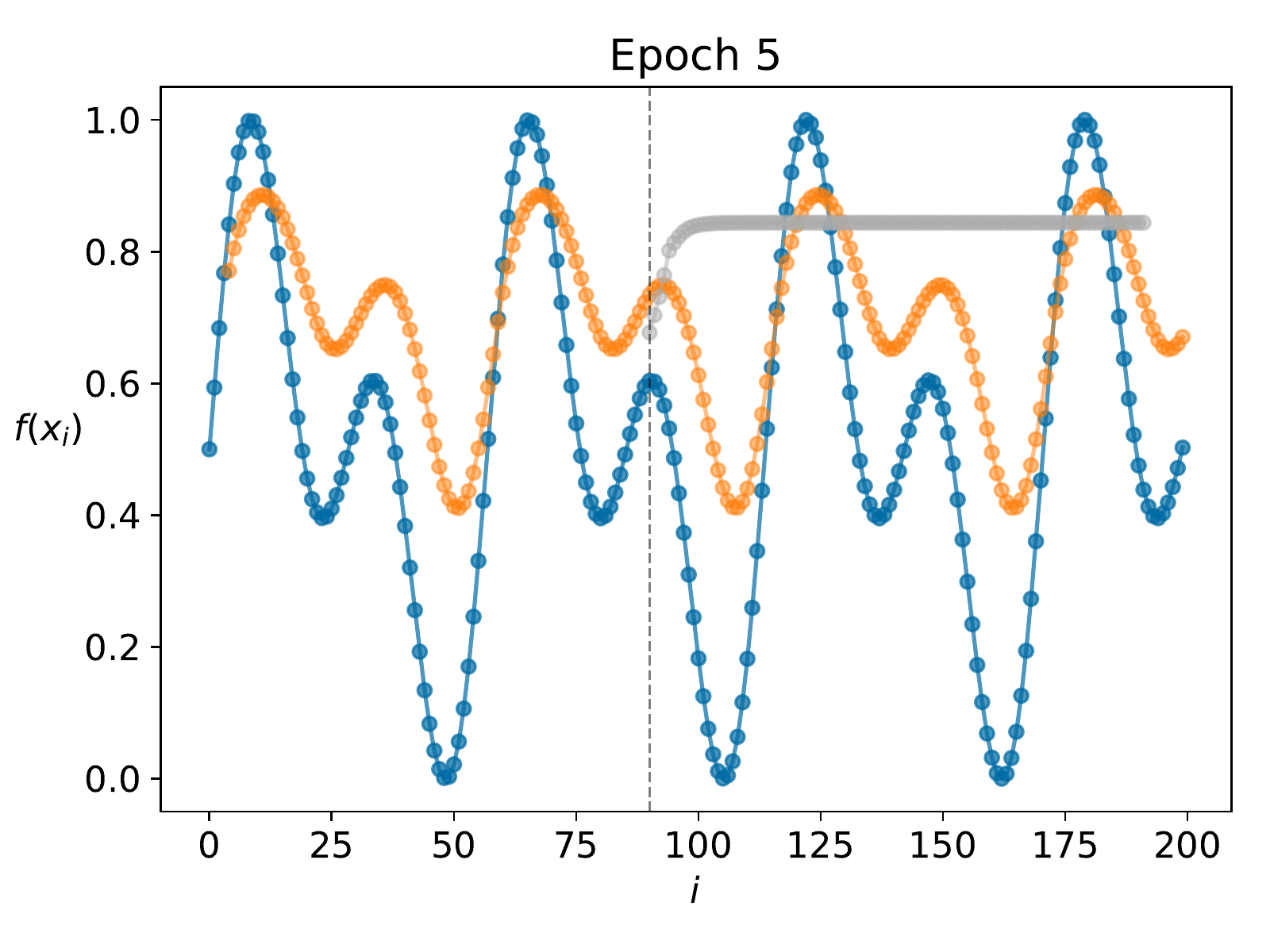}
  \end{subfigure}%
  \begin{subfigure}{0.5\textwidth}
    \includegraphics[width=\textwidth]{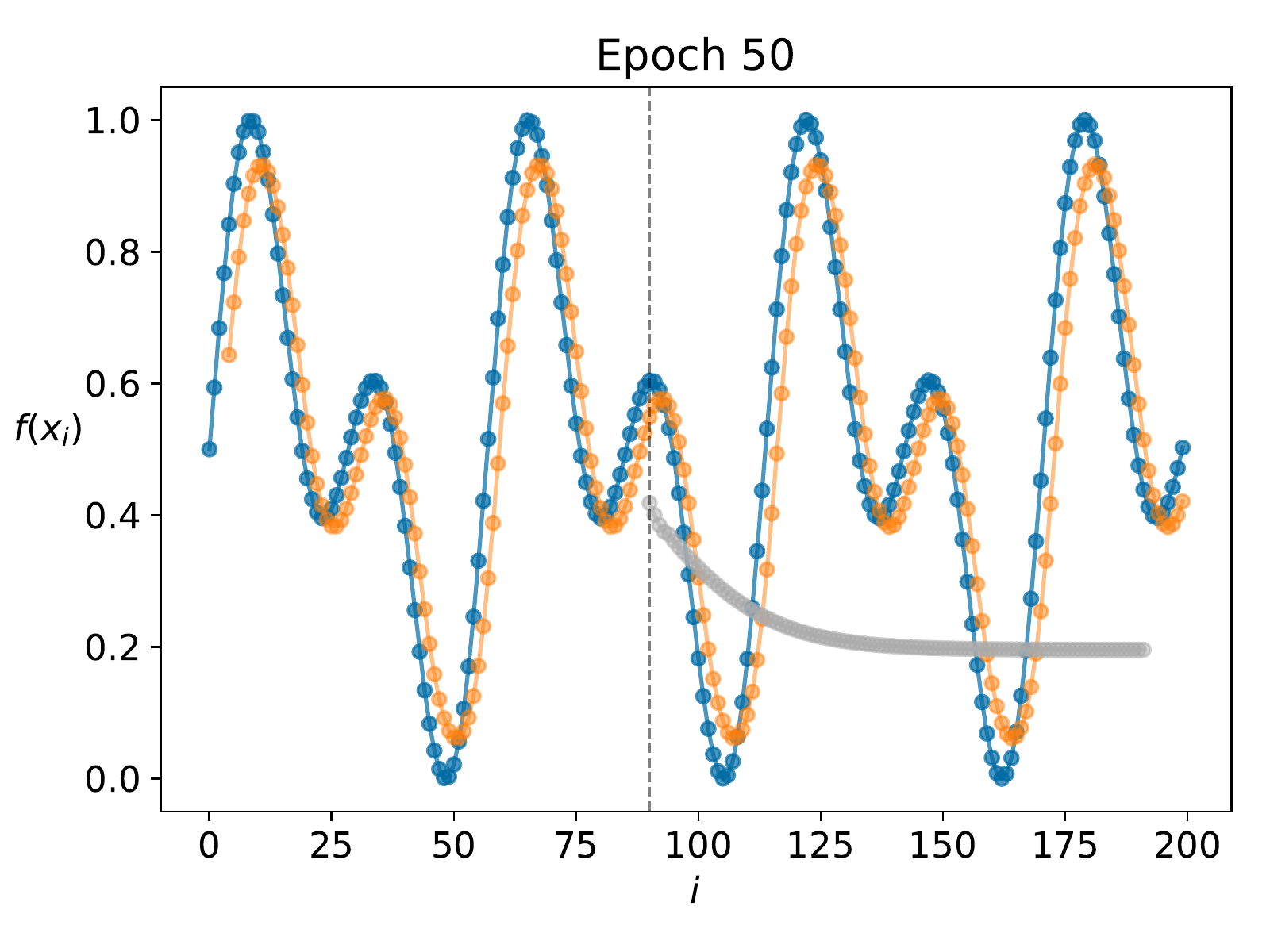}
  \end{subfigure}
    \caption{Progress of training on the data generated with
      function ${\frac{1}{2} \sin(x) + \frac{1}{2} \sin(2x)}$, for
        CV-QRNN (top row) and LSTM networks (bottom row). Blue points
        represent the reference data, orange points are predictions based
        on $T=4$ previous points, and the gray ones -- the forecasted
        values. Vertical dashed line marks the point where the data was
        split for training (left) and testing (right) sequences.}
  \label{fig:sine_wave_2_epochs} 
\end{figure}

\begin{figure}[ht]
  \centering
  \begin{subfigure}{0.5\textwidth}
    \includegraphics[width=\textwidth]{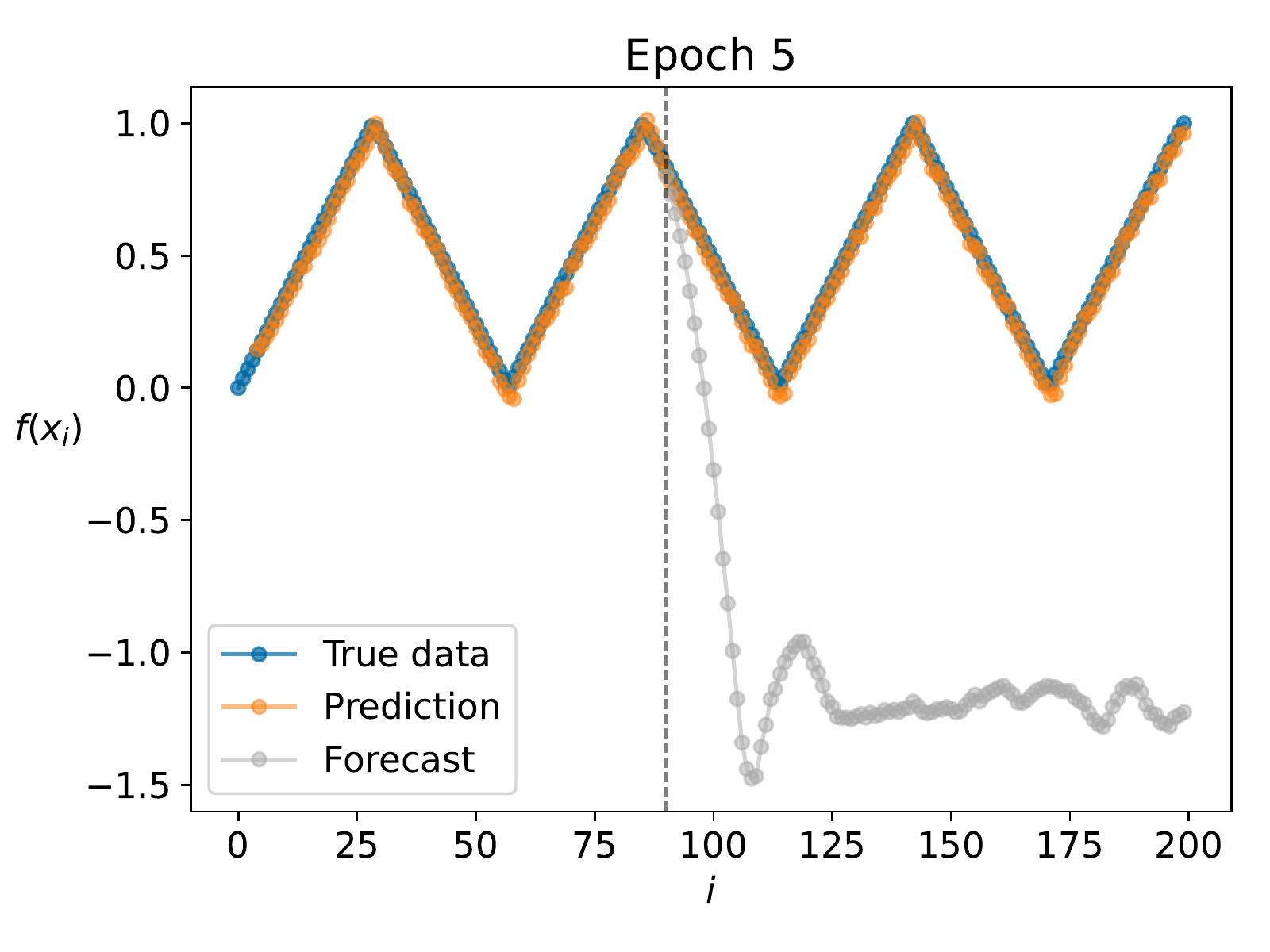}
  \end{subfigure}%
  \begin{subfigure}{0.5\textwidth}
    \includegraphics[width=\textwidth]{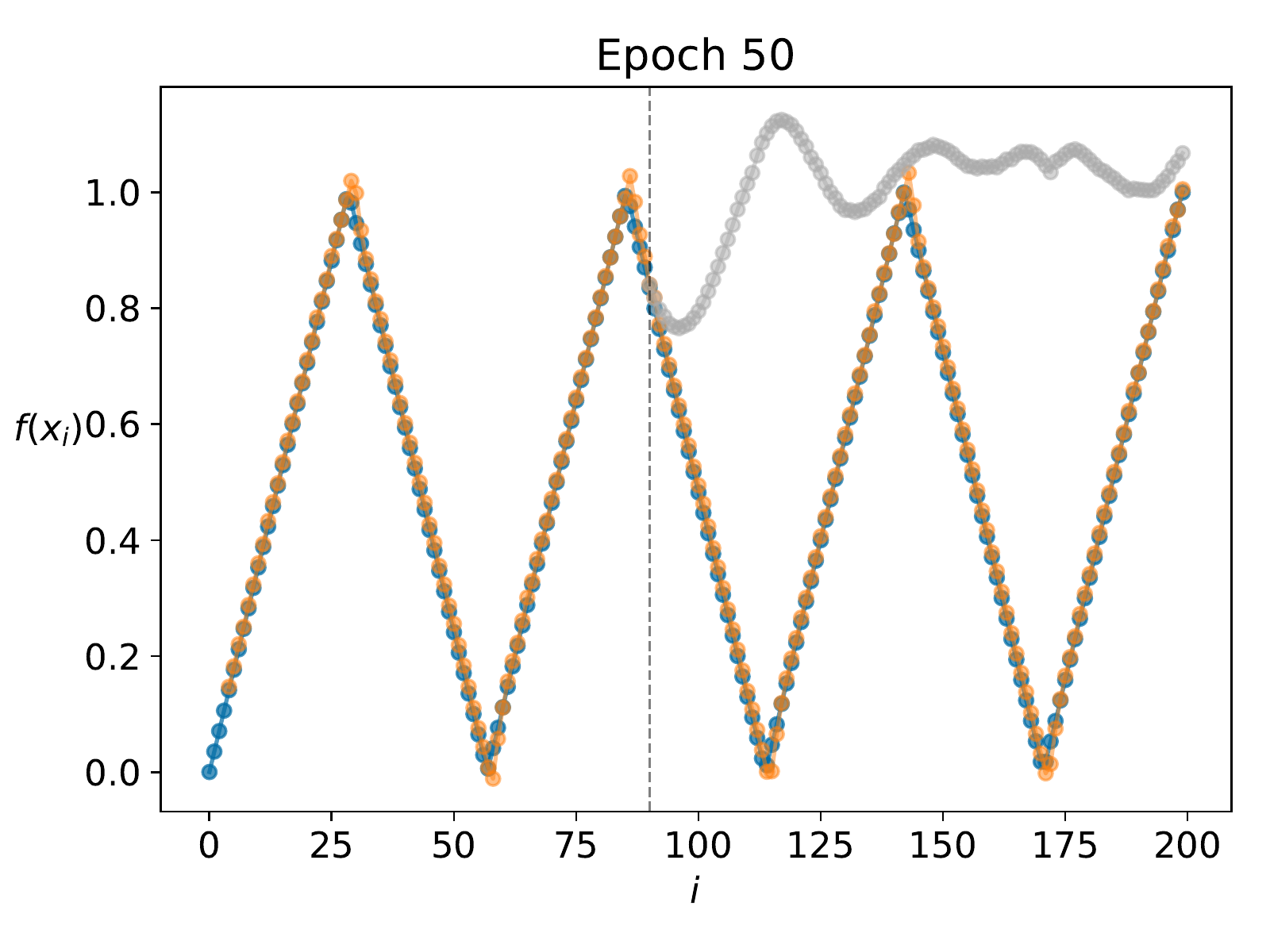}
  \end{subfigure}
  \hfill
  \begin{subfigure}{0.5\textwidth}
    \includegraphics[width=\textwidth]{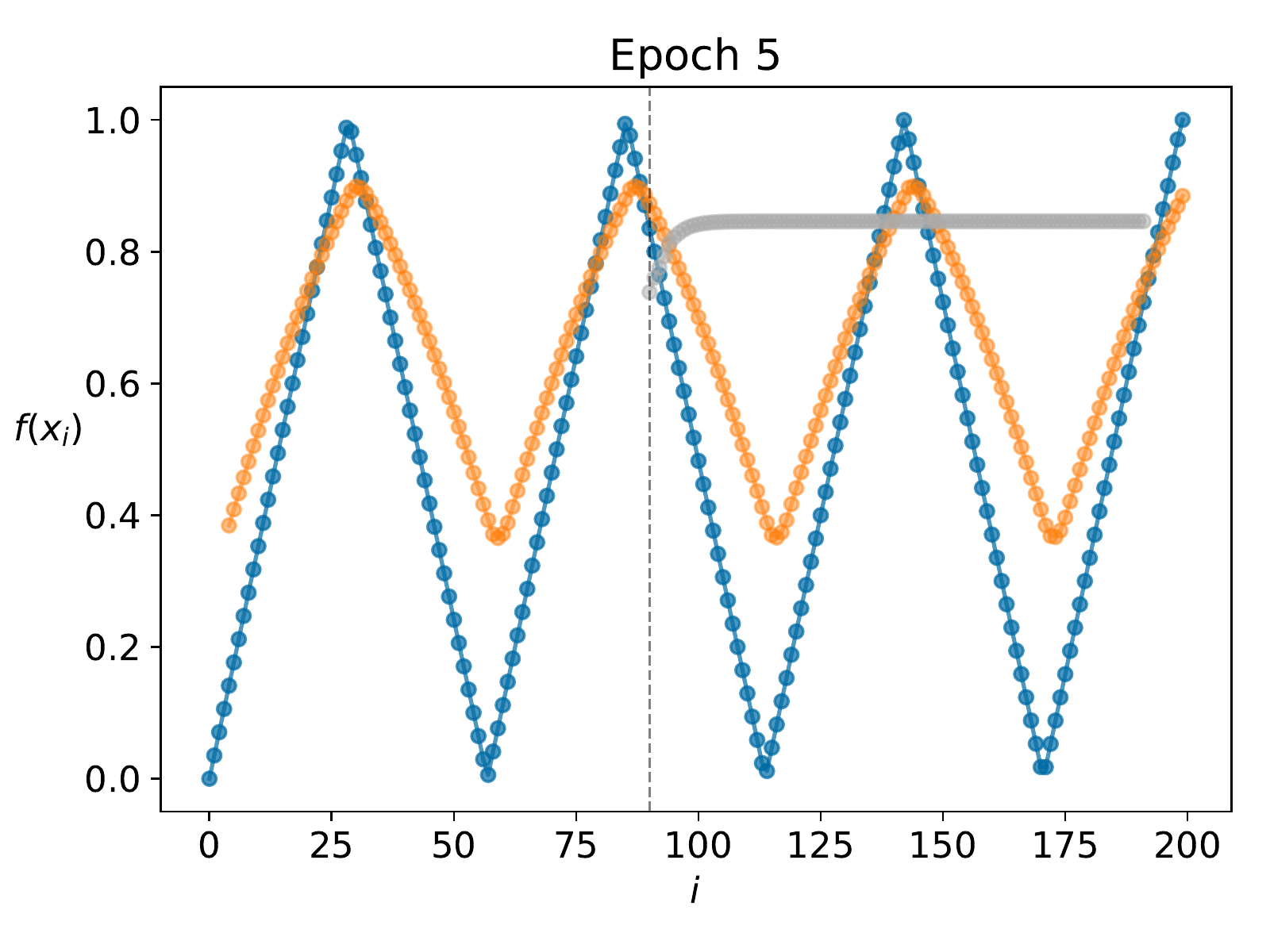}
  \end{subfigure}%
  \begin{subfigure}{0.5\textwidth}
    \includegraphics[width=\textwidth]{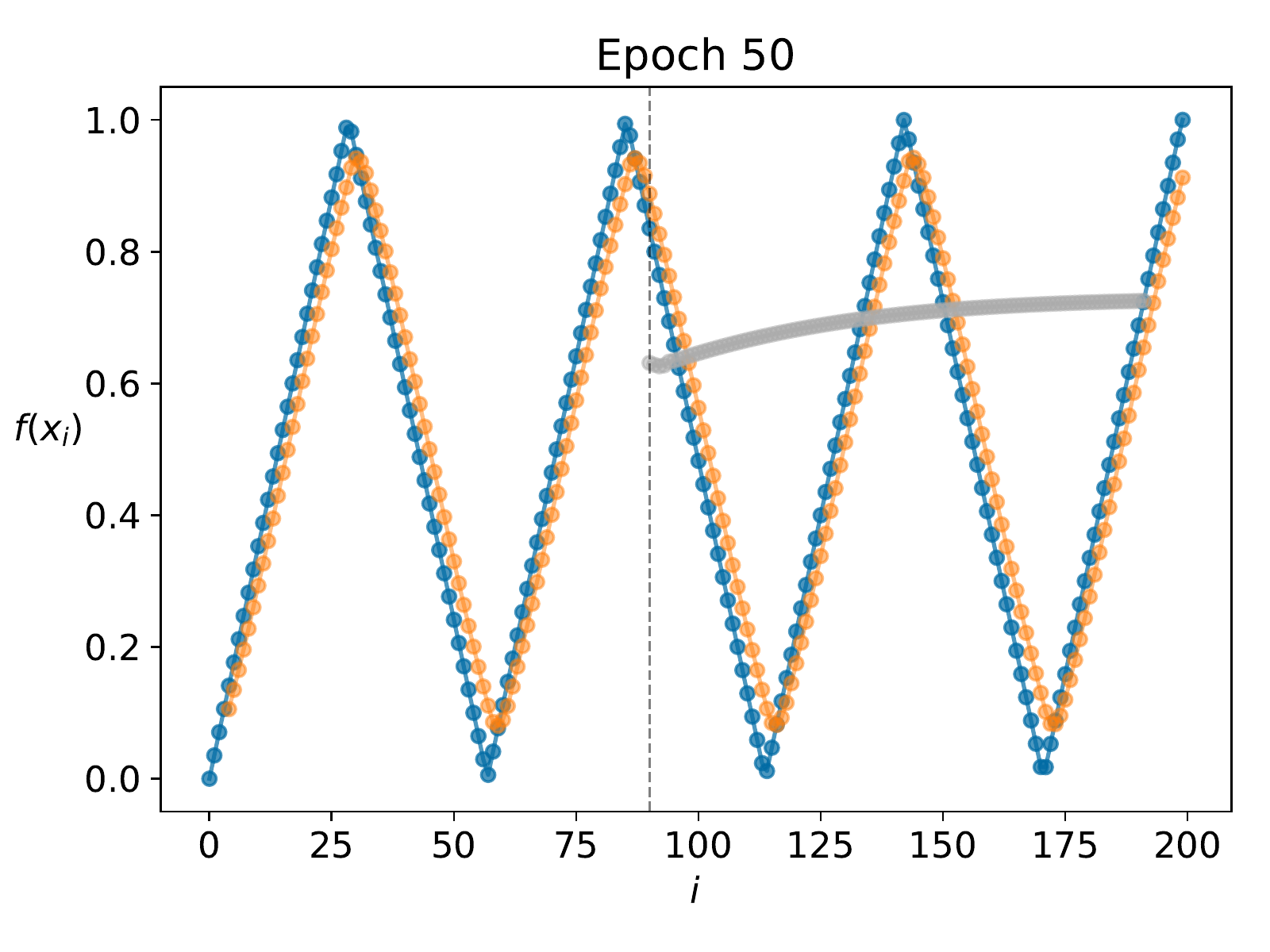}
  \end{subfigure}
  \caption{Progress of training on the data generated with triangle wave
    function, for CV-QRNN (top row) and LSTM networks (bottom row). Blue
    points represent the reference data, orange points are predictions
    based on $T=4$ previous points, and the gray ones -- the forecasted
    values. Vertical dashed line marks the point where the data was split
    for training (left) and testing (right) sequences.}
  \label{fig:triangle_wave_epochs} 
\end{figure}

\begin{figure}[ht]
  \centering
  \begin{subfigure}{0.5\textwidth}
    \includegraphics[width=\textwidth]{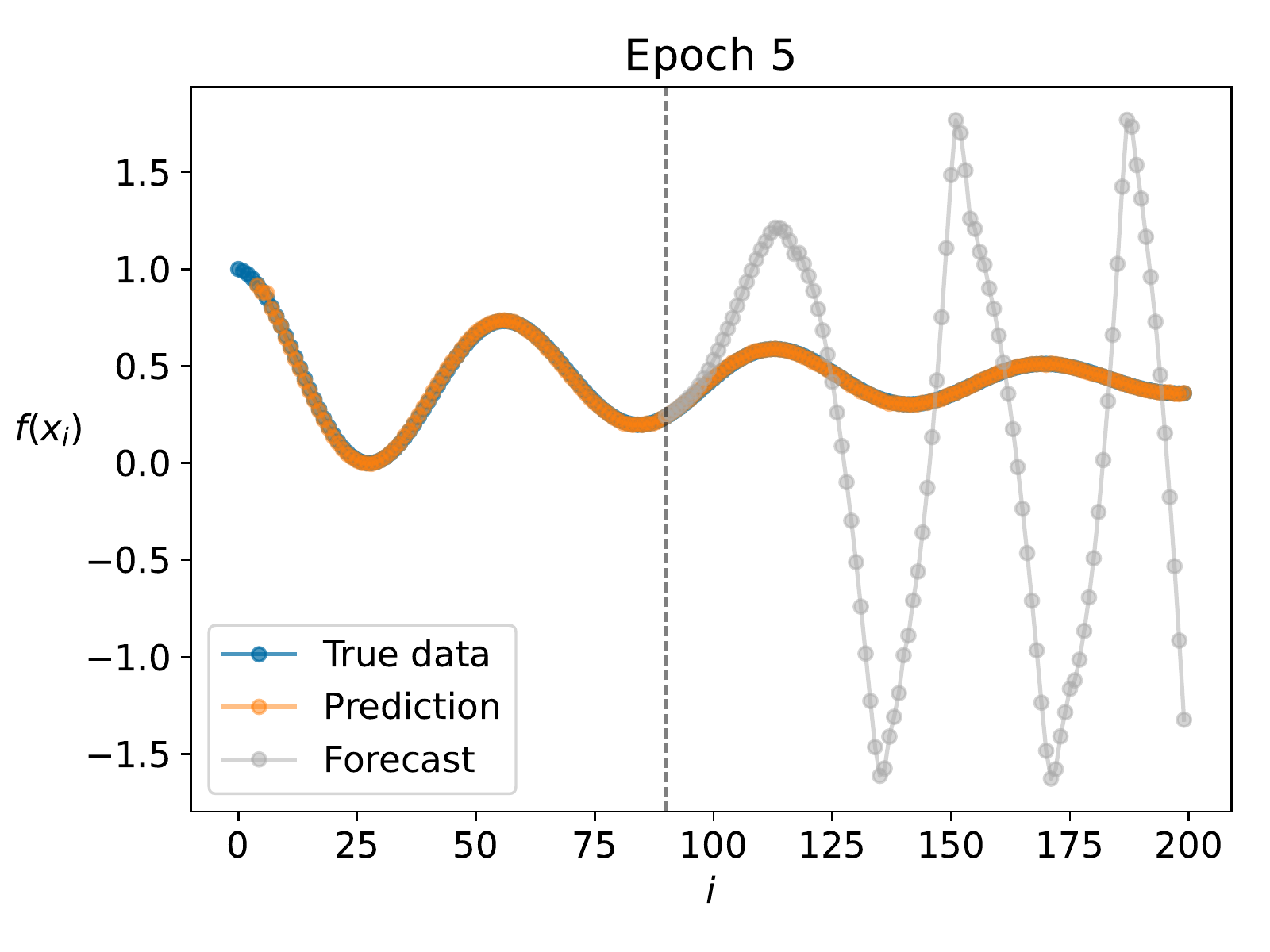}
  \end{subfigure}%
  \begin{subfigure}{0.5\textwidth}
    \includegraphics[width=\textwidth]{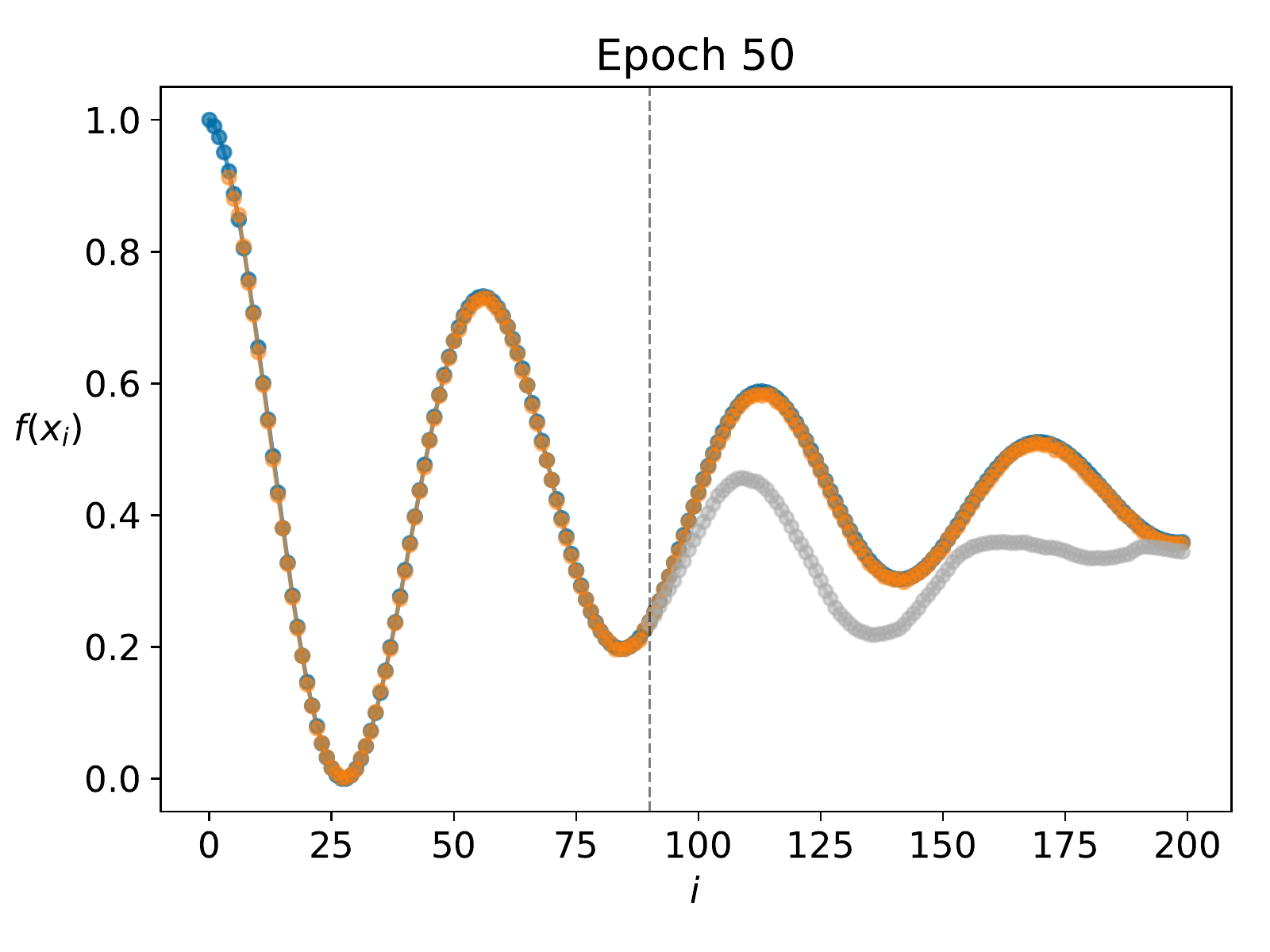}
  \end{subfigure}
  \hfill
  \begin{subfigure}{0.5\textwidth}
    \includegraphics[width=\textwidth]{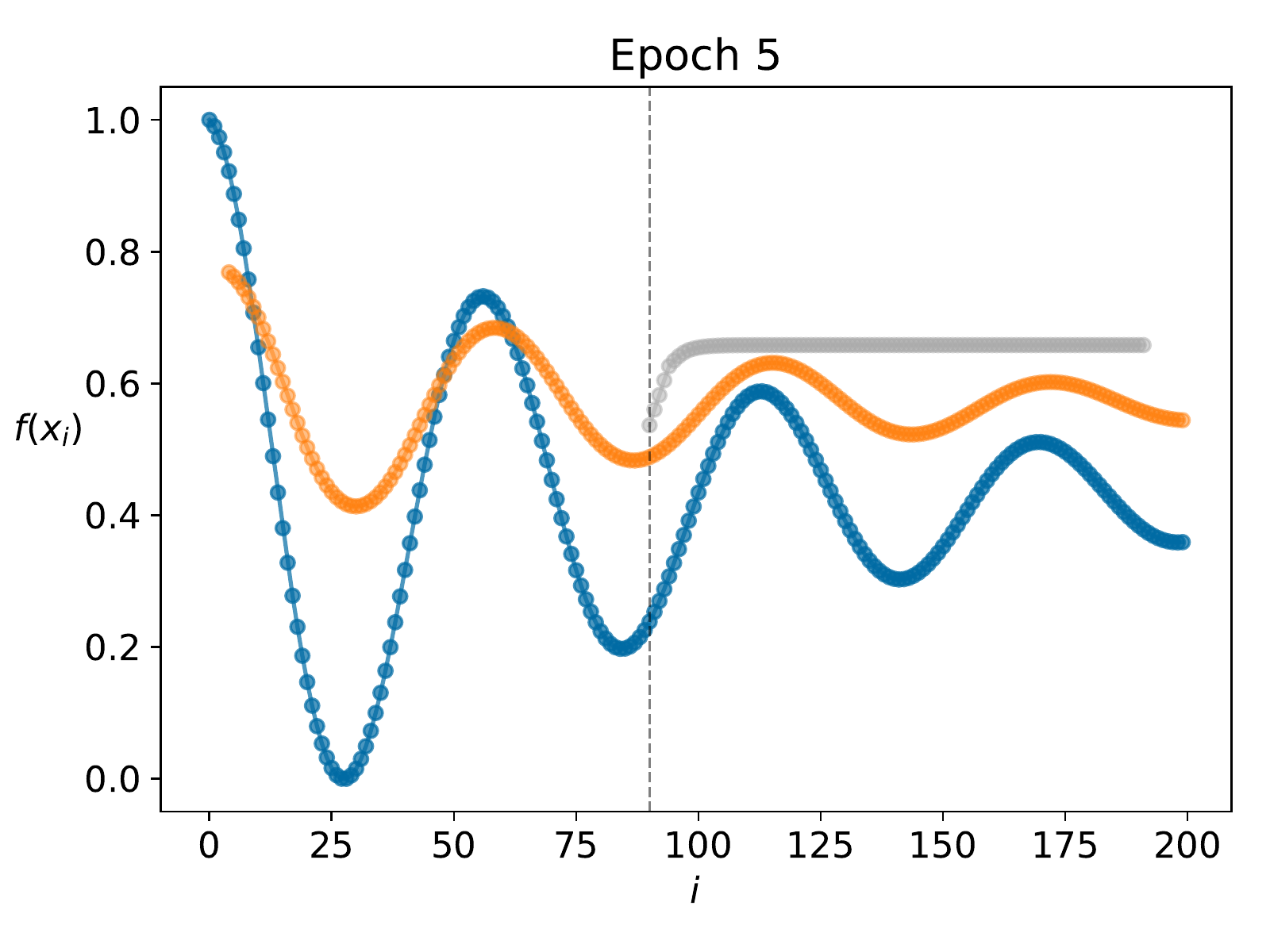}
  \end{subfigure}%
  \begin{subfigure}{0.5\textwidth}
    \includegraphics[width=\textwidth]{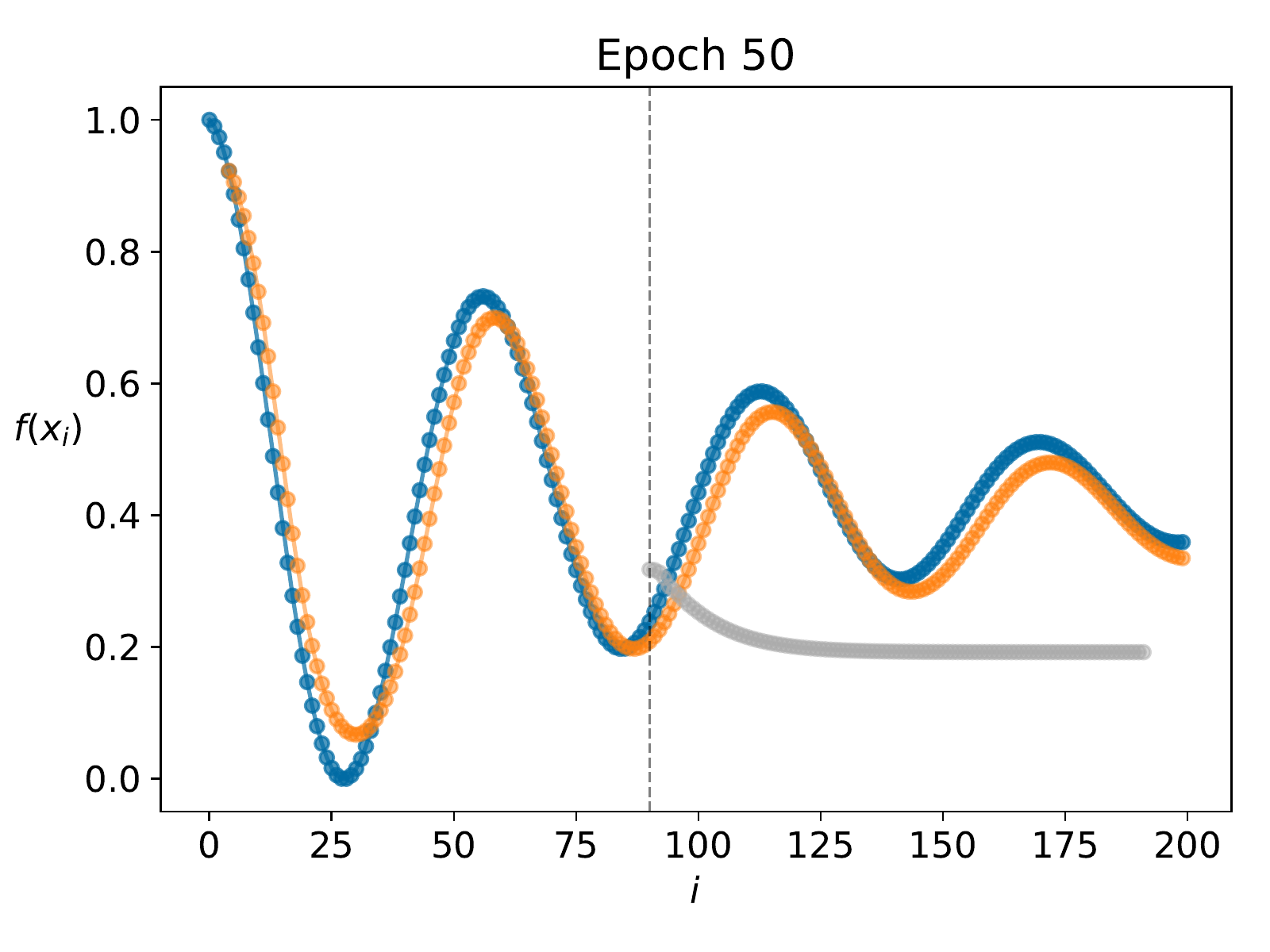}
  \end{subfigure}
  \caption{Progress of training on the data generated with function
    ${\exp(-\frac{x}{10}) \cdot \cos(x)}$ (so called damped
    oscillation), for CV-QRNN (top row) and LSTM networks (bottom row).
    Blue points represent the reference data, orange points are
    predictions based on $T=4$ previous points, and the gray ones --
    the forecasted values. Vertical dashed line marks the point where
    the data was split for training (left) and testing (right)
  sequences.}
  \label{fig:cos_wave_damped_epochs} 
\end{figure}

Here, we present results for other data sets: sine wave
(\cref{fig:sin_loss} and \cref{fig:sine_wave_epochs}), sum of two sine
waves with period $2\pi$ and $\pi$ (\cref{fig:sin_loss_2} and
\cref{fig:sine_wave_2_epochs}), triangle wave (\cref{fig:triangle_loss}
and \cref{fig:triangle_wave_epochs}) and exponentially damped cosine wave
(\cref{fig:cos_loss} and \cref{fig:cos_wave_damped_epochs}). We depict
the cost function during the training of the network in
\cref{fig:loss_rest}. The prediction and forecasting ability is presented
in
\cref{fig:sine_wave_epochs,fig:sine_wave_2_epochs,fig:triangle_wave_epochs,fig:cos_wave_damped_epochs}
for data sets described previously. For these data sets we found similar
results as for the Bessel function, which was described in the main text.

\end{document}